\documentclass[11pt,a4paper,iop,apj]{emulateapj}
\usepackage{epsfig}
\usepackage{color}
\begin{document}
\title{The Araucaria Project. The distance to the Small Magellanic Cloud from late-type eclipsing binaries}
\author{Dariusz Graczyk\altaffilmark{1}, Grzegorz Pietrzy{\'n}ski\altaffilmark{2,1}, Ian B. Thompson\altaffilmark{3}, Wolfgang Gieren\altaffilmark{1,5,6}, Bogumi{\l} Pilecki\altaffilmark{2,1}, Piotr Konorski\altaffilmark{2}, Andrzej Udalski\altaffilmark{2}, Igor Soszy{\'n}ski\altaffilmark{2}, Sandro Villanova\altaffilmark{1}, Marek G{\'o}rski\altaffilmark{2}, Ksenia Suchomska\altaffilmark{2}, Paulina Karczmarek\altaffilmark{2}, Rolf-Peter Kudritzki\altaffilmark{4}, Fabio Bresolin\altaffilmark{4} and Alexandre Gallenne\altaffilmark{1}}
\affil{$^1$Universidad de Concepci{\'o}n, Departamento de Astronomia, Casilla 160-C, Concepci{\'o}n, Chile}
\affil{$^2$Warsaw University Observatory, Al. Ujazdowskie 4, 00-478 Warsaw, Poland }
\affil{$^3$Carnegie Observatories, 813 Santa Barbara Street, Pasadena, CA 911101-1292, USA}
\affil{$^4$Institute for Astronomy, University of Hawaii at Manoa, 2680 Woodlawn Drive, Honolulu, HI 96822, USA}
\affil{$^5$ University Observatory Munich, Scheinerstrasse 1, 81679 Munich, Germany}
\affil{$^6$ Max Planck Institutefor Extraterrestial Physics, Giessenbachstrasse, 85748, Garching, Germany}

\begin{abstract}
We present a distance determination to the Small Magellanic Cloud (SMC)  based on an analysis of four  detached, long period, late type eclipsing binaries discovered by the OGLE Survey. The components of the binaries show negligible intrinsic variability. A consistent set of stellar parameters was derived with  low statistical and systematic uncertainty. The absolute dimensions  of the stars are calculated with a precision of better than 3\%. 

The surface brightness - infrared color relation was used to derive the distance to each binary. The four systems clump around a distance modulus of  $(m\!-\!M)=18.99$ with a dispersion of only 0.05 mag. Combining these results with the distance published by Graczyk et al.~for the eclipsing binary OGLE SMC113.3 4007 we obtain a mean distance modulus to the SMC of 18.965 $\pm$ 0.025 (stat.) $\pm$ 0.048 (syst.) mag. This corresponds to a distance of 62.1 $\pm$ 1.9 kpc, where the error includes both uncertainties. Taking into account other recent published determinations of the SMC distance we calculated the distance modulus difference between the SMC and the LMC equal to 0.458 $\pm$ 0.068 mag.  Finally we advocate $\mu_{SMC}=18.95\pm0.07$ as a new "canonical" value of the distance modulus to this galaxy.
\end{abstract} 

\section{Introduction}

The Small Magellanic Cloud (SMC) is a dwarf irregular galaxy connected  by a hydrogen gas and stellar bridge to the Large Magellanic Cloud (LMC).  Both are nearby satellites of our Milky Way galaxy and because of this  proximity they are ideal environments within which to study stellar populations and to calibrate many different stellar standard candles. The SMC is more  metal-poor than the LMC, and this provides an important opportunity to check on  the metallicity dependencies of the different standard candles used  in distance scale work. This makes the SMC very important in establishing  a precise distance scale despite its complicated geometrical  structure and larger extension in the line-of-sight compared to the LMC.  For these reasons, the SMC has  played an important role in our  Araucaria Project \citep{gie05} whose main goal is to produce a significantly improved calibration of the extragalactic distance scale. 

For almost a decade we have been observing late-type eclipsing binary systems in both Magellanic Clouds, systems which have a particularly strong potential to derive precise distances these galaxies \citep{pie09,gra12}. Recent observations have proven this potential: analysis of eight binaries in the LMC  has led  to a distance to the LMC accurate to 2.2\%, more accurate than any other previous determination of this fundamental parameter \citep{pie13}. Here we extend this work to a measurement of the distance to the SMC using the same method, leading in addition to a better knowledge of the geometrical structure of this galaxy. From our studies of late-type detached  eclipsing binary systems in the SMC and LMC we also expect to  improve on  the determination of the astrophysical parameters of red giant stars and  Cepheid variables \citep{pie10,pil13}. 

The advantage of using detached late type systems over those containing early type stars comes from utilizing well determined empirical surface brightness relation (e.g.~the Barnes-Evans relation in the optical) derived for late type stars from interferometric measurements (e.g.~\cite{bar76}, \cite{ker04}, \cite{ben05}). The main points are well summarized by \cite{lac77}: "The distances are derived without assumptions about luminosity class or effective temperature, and are ultimately based only on geometrical factors. That means that the only source for systematic error should be from the Barnes-Evans calibration  and this is the place where we expect the most significant improvement will occur. Another wirtue is that for stars later than A0 the Barnes-Evans distance is essentially independent of reddening."

The paper is organized as follows. In Section 2 we describe photometric and spectroscopic observations, Section 3 gives details of our method, Section 4 contains a description of model solutions for individual binaries. In section 5 we present a distance determination to the SMC and Section 6 is devoted to a discussion of our results. 

\section{Observations and Data Reduction}
\label{obs}
\begin{figure}
\includegraphics[angle=0,scale=.575]{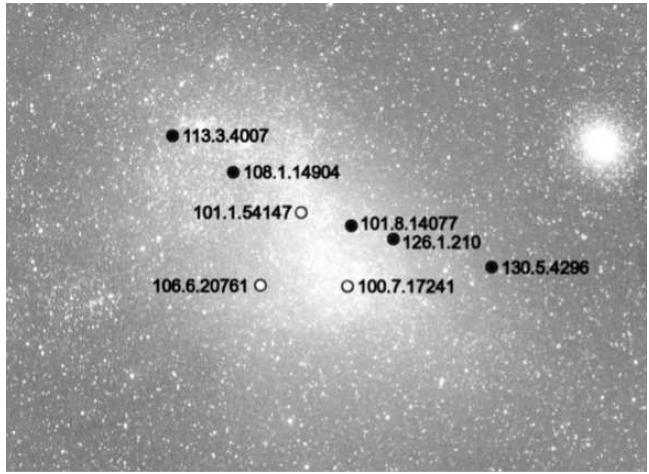}
\caption{ The positions of the eclipsing binaries within the SMC. North is up, east on the left. The filled circles denote four systems from this paper and one system previously analyzed by \cite{gra12} on upper left - SMC113.3 4007. The open circles represent three additional systems for which the analysis is not  yet finished. \label{fig0}}
\end{figure}

Basic data on our four eclipsing binary systems are given in Table~\ref{tbl:1}. Their position within the body of the SMC is presented in Fig.~\ref{fig0}. With the exception of SMC130.5 4296 our targets were discovered by \cite{uda98} and confirmed to be eclipsing binaries by \cite{wyr04}. Optical photometry in the Johnson-Cousins filters  was obtained with the Warsaw 1.3 m telescope at Las Campanas Observatory in the course of the second, third and fourth phase of the OGLE project \citep{uda97,uda03,sos12}.  We secured 5314 I-band measurements (over 1000 per system) and 391 V-band measurements (over 70 per system) for five eclipsing binaries. Because of the long orbital periods consecutive epochs were taken usually on different nights. The time span of the I-band observations is 5684 days. The raw data were reduced with the image-subtraction technique \citep{woz00,uda03} and instrumental magnitudes were calibrated onto the standard system using Landolt standards.  

Near-infrared  photometry was collected with the ESO NTT telescope on La Silla, equipped with the SOFI camera (PI:  Pietrzy{\'n}ski). The setup of the instrument, reduction and calibration of the data onto the UKIRT system were essentially the same as described in  \cite{pie09}. We collected at least 10 epochs of infrared photometry for every our target outside eclipses. The transformation of our photometry onto the Johnson system was done using the equations given by \cite{car01} and \cite{bes88}. The whole multiband photometry is presented in Table \ref{tbl:phot}. 

High resolution echelle spectra were collected with the Clay 6.5 m telescope at Las Campanas, equipped with the MIKE spectrograph and with the 3.6 m telescope in La Silla, equipped with the HARPS spectrograph. We used a $5\times0.7$ arcsec slit with  MIKE giving a resolution of about 40000.  In the case of HARPS we used the EGGS mode giving a resolution of about R$\sim$80000. The typical S/N at $\sim$5300 \AA~was about 20 and 6 for the MIKE-Red part and HARPS spectra, respectively, and about 12 at $\sim$4700 \AA~for the MIKE-Blue part spectra. 

In order to determine radial velocities of the system's components we employed the Broadening Function (BF) formalism introduced by \citet{ruc92,ruc99}. Radial velocities were derived using the RaveSpan software \citep{pil12} using numerous metallic lines in the wavelength regions 4125-4230, 4245-4320, 4350-4840, 4880-5000, 5350-5850, 5920-6250, 6300-6390, 6600-6800 \AA. As templates we used synthetic spectra with [Fe/H]=$-0.5$ from a library computed by \cite{col05}. The templates were chosen to match the atmospheric properties of the stars in a grid of $T_{\rm eff}$ and $\log g$. The resulting velocities are presented in Table \ref{tbl:spec}. We calculated instrumental shifts taking into account residual radial velocities from our best models. They are: MIKE-RED$-$MIKE-BLUE$= +3\pm85$ m s$^{-1}$ and MIKE-BLUE$-$HARPS$= +48\pm114$ m s$^{-1}$. We concluded that there is no significant systematic shift in radial velocities from spectra taken by HARPS and MIKE spectrographs. In the analysis we included separately radial velocities from both the blue and red sides of MIKE, effectively including the MIKE velocities at  twice the weight of the HARPS velocities. The reason is that the  MIKE spectra usually have significantly better S/N compared to the HARPS spectra. 
 
\begin{deluxetable*}{@{}lcclcccc}
\tabletypesize{\scriptsize}
\tablecaption{The target stars \label{tbl:1}}
\tablewidth{0pt}
\tablehead{
\colhead{OGLE ID} & \colhead{RA} & \colhead{Dec} & \colhead{$V$} & \colhead{$V-I$} & \colhead{$V-K$} & \colhead{$J-K$} & \colhead{$P_{\rm obs}$}  \\
\colhead{} & \colhead{h:m:s} & \colhead{deg:m:s} & \colhead{mag} & \colhead{mag} & \colhead{mag} & \colhead{mag} & \colhead{d} 
}
\startdata
 SMC101.8 14077 &  00:48:22.70 &  -72:48:48.6 &  17.177  &  0.935 & 2.099 & 0.489 & 102.90\\
 SMC108.1 14904 &  01:00:18.10 &  -72:24:07.1 &  15.205  &  0.963 & 2.194 & 0.556 &185.22 \\
 SMC126.1 210      &  00:44:02.68 &  -72:54:22.5 &  16.771  &  1.249 & 2.941 & 0.765 & 635.00 \\
 SMC130.5 4296   &  00:33:47.90  &  -73:04:28.0 &  16.783  &  1.207 & 2.788 & 0.754 & 120.47 
\enddata
\tablecomments{Given are coordinates, observed magnitudes, colors and orbital periods.  Identification numbers are from the OGLE-III database \citep{uda03}.}
\end{deluxetable*}

\begin{deluxetable}{@{}lcrcc}
\tabletypesize{\scriptsize}
\tablecaption{The photometry of target systems \label{tbl:phot}}
\tablewidth{0pt}
\tablehead{
\colhead{Name} & \colhead{Band} & \colhead{HJD} & \colhead{Mag} & \colhead{Err}  \\
\colhead{} & \colhead{} & \colhead{HJD-2450000} & \colhead{mag} & \colhead{mag} 
}
\startdata
smc101 & I & 621.80276 & 16.228 & 0.010 \\
smc101 & I & 622.87983 & 16.235 & 0.008 \\
smc101 & I & 624.90639 & 16.252 & 0.008 \\
smc101 & I & 625.89383 & 16.265 & 0.008 \\
smc101 & I & 626.86536 & 16.258 & 0.008 \\
smc101 & I & 627.85158 & 16.252 & 0.008 
\enddata
\tablecomments{This table is avalaible entirety in machine-readable format in the online journal. A portion is shown here for guidance regarding its form and content.}
\end{deluxetable}

\begin{deluxetable*}{@{}lrccccl}
\tabletypesize{\scriptsize}
\tablecaption{The radial velocities measurements \label{tbl:spec}}
\tablewidth{0pt}
\tablehead{
\colhead{Name} & \colhead{HJD} & \colhead{RV1} & \colhead{Err1} & \colhead{RV2} & \colhead{Err2} &\colhead{Instrument}  \\
\colhead{} & \colhead{HJD-2450000} & \colhead{km s$^{-1}$} & \colhead{km s$^{-1}$} &  \colhead{km s$^{-1}$} & \colhead{km s$^{-1}$} & \colhead{}
}
\startdata
smc101 & 4328.78143 & 224.764 & 0.300 & 153.453 & 0.300 & MIKE-Red \\ 
smc101 & 4328.78161 & 224.407 & 0.300 & 153.465 & 0.300 & MIKE-Blue \\
smc101 & 4329.81804 & 225.409 & 0.300 & 152.038 & 0.300 & MIKE-Blue \\
smc101 & 4329.81804 & 225.577 & 0.300 & 151.729 & 0.300 & MIKE-Red \\ 
smc101 & 4395.57940 & 150.271 & 0.300 & 221.691 & 0.300 & HARPS \\
smc101 & 4424.58869 & 212.609 & 0.300 & 163.422 & 0.300 & MIKE-Blue
\enddata
\tablecomments{This table is avalaible entirety in machine-readable format in the online journal. A portion is shown here for guidance regarding its form and content.}
\end{deluxetable*}

\section{Modeling approach}
The light and radial velocity curves were analyzed using the Wilson-Devinney (hereafter WD) program, version 2007 \citep{wil71,wil79,wil90,van07}. We follow a methodology of analysis as  described in Graczyk et al. (2012) and the Supplementary Information section of Pietrzy{\'n}ski et al. (2013). Here we  give some additional details of our approach. The analysis can be divided into "preliminary" and "final" steps. During the "preliminary" analysis we use all light curves, radial velocities, infrared colors and a first estimation of the reddening to set initial constraints on model parameters and to find a preliminary solution. This solution is used to disentangle the spectra. In the "final" analysis we add information from spectral disentangling to improve input model parameters and derive the final solution. 

\subsection{Parameter  Choice}
\label{par}

\begin{deluxetable}{lcc}
\tabletypesize{\scriptsize}
\tablecaption{The intensity ratios \label{tbl-2}}
\tablewidth{0pt}
\tablehead{
\colhead{ID} & \colhead{Blue ($\sim 4500$ \AA)} & \colhead{Red ($\sim 6000$ \AA)} 
}
\startdata
SMC101.8 14077 & $1.37 \pm 0.05$ & $1.61 \pm 0.05$ \\
SMC108.1 14904 & $0.66 \pm 0.02$ & $1.00 \pm 0.02$ \\
SMC126.1 210 & $0.95 \pm 0.02$ & $0.91  \pm 0.02$\\
SMC130.5 4296 & $2.1 \pm 0.1$ & $2.6 \pm 0.3$
\enddata
\tablecomments{The relative strengths of the absorption lines of the secondary  with respect to those of the primary.}
\end{deluxetable}

\begin{deluxetable*}{@{}lcccccccc}
\tabletypesize{\scriptsize}
\tablecaption{Atmospheric parameters\label{tbl-4}}
\tablewidth{0pt}
\tablehead{
\colhead{Property} &\multicolumn{2}{c}{{SMC101-8-14077}}& \multicolumn{2}{c}{{SMC108-1-14904}}& \multicolumn{2}{c}{{SMC126-1-210}}& \multicolumn{2}{c}{{SMC130-5-4296}} \\
&\colhead{Primary} & \colhead{Secondary} &\colhead{Primary} & \colhead{Secondary} &\colhead{Primary} & \colhead{Secondary} &\colhead{Primary}&\colhead{Secondary} 
}
\startdata
 $T_{\rm eff}$	(K) 		& - & 5080 &5350&5200&4500&4540&5050& 4420\\
 $\log{g}$	(cgs)		& - & 1.20   &0.40 &2.20 &1.45 &1.10 &2.45  &1.75\\
$[$Fe/H$]$ 			& - &-1.01 &-0.95&-0.64&-0.94&-0.79&-0.77&-0.99\\
$v_t$ (km $s^{-1}$)	& - &1.95   &3.50 &2.40 &1.60 &1.79&1.67  &2.08
 \enddata
 \end{deluxetable*}
 
The orbital periods were calculated with the string-length method (Lafler \& Kinman 1965, Clarke 2002). The phased light curves were inspected visually to set a preliminary epoch of the primary minimum and to verify the photometric stability of the systems - a lack of spots, flares, the O'Connell effect (O'Connell 1951) - at the precision level of our photometry. 

The effective temperature of the primary component of each system (eclipsed during minimum at orbital phase 0) was estimated as follows. We determined the interstellar extinction in the direction of our targets using \cite{has11} reddening maps - see Section~\ref{red}. Average out-of-eclipse magnitudes were calculated from all observations taken outside photometric minima. In the case of the two systems showing  proximity effects between eclipses we used magnitudes only around both quadratures. All out-of-eclipse V-band magnitudes and ($V\!-\!I$) colors are presented in Table~\ref{tbl:1}. These magnitudes were dereddened using the interstellar extinction law given by Cardelli et al. (1989) and O'Donnell (1994) and assuming R$_V=3.1$. 

To set the effective temperature scale of each system we ran the WD code initially assuming a temperature of $T_1=5000$ K for the primary and [Fe/H]$\,=-0.5$. The resulting  luminosity ratios in the V, I and K bands were combined with dereddened magnitudes  to calculate first guesses for the intrinsic ($V\!-\!I$) and ($V\!-\!K$) colors of both components. Then we utilized several calibrations between colors and effective temperatures \citep{ben98,alo99,hou00,ram05,mas06,gon09,cas10,wor11} to set new temperatures for the primaries. We iterated  these steps until new derived temperatures changed by  less than 10 K. The resulting effective temperatures were used in the "preliminary" solution. 

We estimated rotational velocities of the components from a Broadening Function analysis. Rotation in all cases is consistent with synchronous rotation, thus we set the rotation parameter $F=1.0$ for both components.  The albedo parameter was set to 0.5 and the gravity brightening to 0.32, both values appropriate for a cool, convective atmosphere. The limb darkening coefficients were calculated internally by WD code (setting LD=$-2$) according to the logarithmic law \citep{kli70} during each iteration of the Differential Correction (DC) subroutine using tabulated data computed by \cite{van93}.         

As free parameters of the WD model we chose the orbital period $P_{\rm obs}$, the semimajor axis $a$, the orbital eccentricity $e$, the argument of periastron $\omega$, the epoch of the primary spectroscopic conjunction $T_0$, the systemic radial velocity $\gamma$, the orbital inclination $i$ , the secondary star average surface temperature $T_2$, the modified surface potential of both components ($\Omega_1$ and $\Omega_2$), the mass ratio $q$, and the relative monochromatic luminosity of the primary star in two bands ($L1_V$ and $L1_I$). 

\subsection{Fitting Procedure}  
\label{fit}
For each star we simultaneously fitted the I-band and V-band light curves  and the two radial velocity curves using the DC subroutine from the WD code. The detached configuration (Mode 2) was chosen during all the analyses and a simple reflection treatment (MREF=1, NREF=1) was employed. A stellar atmosphere formulation was selected for both stars (IFAT=1). Level dependent weighting was applied (NOISE=1) and curve dependent weightings (SIGMA) were calculated after each iteration. The grid size was set to N=40 on both components. We did not break the adjustable parameters set into subsets, but instead we adjusted all free parameters at each iteration.  

For each system we used information obtained from spectroscopy to derive the relative intensity of absorption lines of both components from the MIKE spectra. We treat these as  approximate light ratios in the blue and red part of the optical spectra. These light ratios are reported in Table~\ref{tbl-2}. We do not use them to fix light ratios in our models, however, but  use them to rule out alias solutions and to verify if the light curve solution is consistent with the spectroscopy.

 \subsection{Spectral disentangling}
 \label{specdi}
To constrain the important input parameters of the eclipsing binary components we decided to disentangle the spectra of  the individual components and carry out an abundance analysis. The most important parameter is the effective temperature of each component which scales the luminosity ratio of the components and the limb darkening coefficients. Also, by calculating the intrinsic ($V\!-\!I$) colors from the derived temperatures, we can estimate the reddening of each system.    
 
We used the prescription given by Gonzales \& Levato (2005) to disentangle individual spectra of the binary components. The method works in the real wavelength domain. It requires  rather high signal to noise ratio spectra to work properly and thus we were restricted to using only the red MIKE spectra. The red MIKE spectra were shifted and stacked using the radial velocities found by using the Broadening Function as  described in the previous section. No initial secondary spectrum was used in the disentangling. The range of wavelength used was from 4960 \AA ~to 6800 \AA ~where numerous FeI and FeII lines are present.  We emphasise  that we do not use the disentangled spectra to derive new radial velocities because they always give higher dispersion than synthetic templates. The reason is that the relatively low S/N ratio of our disentangled spectra reduces the accuracy of radial velocity measurements. 

The disentangled spectra have to be renormalized in order to account for the companion's continuum which dilutes the depth of the absorption lines. If the light ratio of the secondary to the primary component at a given wavelength is $k(\lambda)=L_{21}(\lambda)$ then we subtract a value $k/(1+k)$ from the disentangled spectrum of the primary (corresponding to the secondary flux) and we again normalize the final spectrum. In the case of the secondary spectrum we subtract $1/(1+k)$. Here two things must be emphasized: 1) the choice of a proper normalization level before disentangling and 2) the choice of a proper model of a system to calculate an appropriate  light ratio function $L_{21}(\lambda)$. Regarding the first point, all spectra have to be normalized and for high S/N spectra the level of normalization is simply the mode value. In the case of lower S/N spectra like those obtained for our binaries in the SMC, the level of normalization is a bit ambiguous and usually we set it to a value of $\sim$90 percentile. 

Regarding the second point, we must be aware that there is a large number of possible model solutions to the observed light and radial velocity curves. A model which is the best i.e. it minimizes the residua of synthetic fits does not always predict light ratios in agreement with spectroscopic light ratios. This is because our data have limited S/N ratio and because an optical depth were absorption lines are formed is usually different from an optical depth at which we observe stellar photosphere. If we use an "improper" model for a system to predict the spectroscopic light ratios, the resulting  absorption line depths in the disentangled spectra are invalid. An easy way to notice an inappropriate model is to check if the absorption line bottoms in the renormalized spectrum of the primary or the secondary are below zero, which is a level signifying an unphysical model. Another check is based on a comparison of equivalent widths of the same absorption lines in both components. Usually the hotter component should have lines with smaller equaivalent widths. However such a comparison is not unique because components of the same binary may have different abundances \citep{fol10}. Finally, we can calculate the equivalent width ratios for the components for the same lines and determine that this ratio is not wavelength dependent.  We expect that for realistic models, this ratio is approximately constant within the spectral range of disentangling.        

\subsection{Atmospheric parameters analysis}
\label{atmo}

\begin{deluxetable*}{lccccccc}
\tabletypesize{\scriptsize}
\tablecaption{Reddenings from the  Na I D1 line \label{tbl:redd}}
\tablewidth{0pt}
\tablehead{\colhead{ID} &\multicolumn{3}{c}{{Galactic component }} &\multicolumn{3}{c}{{SMC component}} & \colhead{Total} \\
\colhead{} &\colhead{$V_r$} &\colhead{Eqv.} &\colhead{E($B\!-\!V$)} &\colhead{$V_r$} &\colhead{Eqv.} &\colhead{E($B\!-\!V$)}&\colhead{E($B\!-\!V$)}\\
\colhead{} &\colhead{(km s$^{-1}$)} & \colhead{(\AA)} &\colhead{(mag)} &\colhead{(km s$^{-1}$)} & \colhead{(\AA)} &\colhead{(mag)} & \colhead{(mag)}
}
\startdata
SMC101.8  14077 &21&0.155&0.049&148&0.091&0.029&0.078\\
SMC108.1 14904 &10&0.171&0.055&129&0.095&0.030&0.085\\
SMC126.1 210	      &13&0.235&0.084&$-$&$ -$&$-$&0.084\\
SMC130.5 4296   &6&0.118&0.037&107&0.082&0.026&0.063
\enddata
\end{deluxetable*}

Atmospheric parameters and the iron content were obtained from the  equivalent widths (EWs) of the iron spectral lines. See \cite{mar08} for a more detailed explanation of the method we used to measure the EWs and \cite{vil10} for a description of  the line list that was used.  We adopted $\log{\epsilon}(Fe)=7.50$ as the solar iron abundance.

Atmospheric parameters were obtained in the following way.  As a first step,  model atmospheres were calculated using ATLAS9 \citep{kur70}  assuming as initial estimations for $T_{\rm eff}$, $\log g$, and $v_t$  values typical for giant  stars (4800 K, 2.00 dex, 1.80 km s$^{-1}$), and [Fe/H]$=-0.8$ dex as a typical value for young stars in the  SMC \citep{dia10}. 

Then $T_{\rm eff}$, $v_t$, and $\log g$ were adjusted and new atmospheric models calculated  in an interactive way in order to remove trends in excitation potential (EP)  and EWs versus abundance for $T_{\rm eff}$ and $v_t$, respectively, and to satisfy the  ionization equilibrium for $\log g$. FeI and FeII were used for this  purpose. The [Fe/H] value of the model was changed at each iteration according  to the output of the abundance analysis. The local thermodynamic equilibrium  program MOOG \citep{sne73} was used for the abundance analysis. 

The derived atmospheric parameters for the eclipsing binary components  are given in Table~\ref{tbl-4}. For all systems with the exception of SMC101.8 14077 we manage to calculate parameters for both components. For SMC101.8 14077 the iron lines of the primary are too weak to carry out an abundance analysis.  Our measurements of abundance are consistent with each component in a binary having the same abundance. The typical accuracy of the parameters are 75 K, 0.4 dex, 0.15 dex and 0.3 km s$^{-1}$ for $T_{\rm eff}$, $\log g$, [Fe/H] and $v_t$, respectively.

 \subsection{Reddening}
 \label{red}
The interstellar extinction in the direction to each of our target stars was derived in three ways. First,  we utilized Magellanic Cloud reddening maps published by Haschke et al. (2011). We calculate an average E($V\!-\!I$) value from all of the reddening estimates within a 2 arc min radius of our stars and divided this average by a factor of 1.28 to get a ($B\!-\!V$) color excess for each system. We then added $\Delta$E($B\!-\!V$)$\,=0.037$ mag, the mean foreground Galactic reddening in the direction of the SMC as derived from the Schlegel et al.(1998) maps. The second column of Table~\ref{tbl-5} gives the resulting reddening to each eclipsing binary. These were used to set the temperature scale of each system and to derive the "preliminary" solution (see Section~\ref{par}).  
 
\begin{deluxetable}{@{}lcccc}
\tabletypesize{\scriptsize}
\tablecaption{Color Excess E($B\!-\!V$) \label{tbl-5}}
\tablewidth{0pt}
\tablehead{\colhead{ID} &\colhead{Haschke} &\colhead{Na I D1} & \colhead{Atmosphere} &\colhead{Adopted}\\
\colhead{} &\colhead{(mag)} &\colhead{(mag)} & \colhead{(mag)} &\colhead{(mag)}
}
\startdata
SMC101.8 14077 &0.076&0.078&0.046&0.067\\
SMC108.1 14904 &0.087&0.085&0.107&0.093\\
SMC126.1 210	      &0.060&0.084&0.097&0.080\\
SMC130.5 4296   &0.072&0.063&0.101&0.079
\enddata
\end{deluxetable}
 
 \begin{figure}
\includegraphics[angle=0,scale=.50]{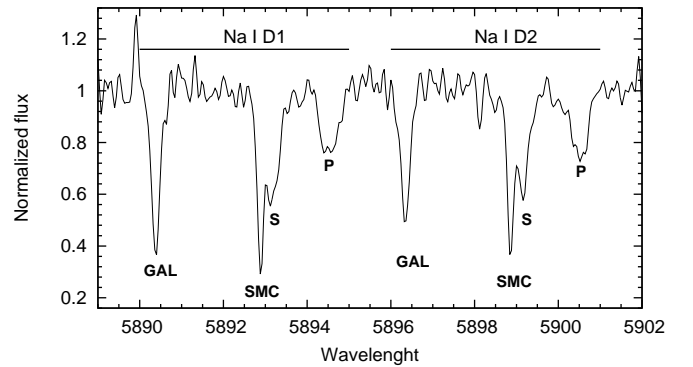}
\caption{The MIKE spectrum of SMC101.8 14077 in the region of the Na I lines (5890.0, 5895.6 \AA) taken close to the third orbital quadrature. Narrow absorption lines arise from galactic (GAL) and  SMC (SMC) interstellar matter. The stellar lines of the primary (P) and secondary (S) are also denoted.  \label{fig:redd}}
\end{figure}

The second method is based on a calibration of the equivalent width of the interstellar Na I D1 line (5890.0 \AA) and reddening (Munari \& Zwitter 1997). The calibration works best for relatively small values of reddening, E($B\!-\!V$)$\,<0.4$ mag. Figure~\ref{fig:redd} presents the spectrum of SMC101.8 14077 around the sodium doublet. Narrow interstellar absorption lines from the Galaxy and the SMC can be identified together with wider stellar absorption from both stars. We separately measured the equivalent widths of both interstellar components and derived the reddening for each.  The results for all our stars is presented in Table~\ref{tbl:redd}. In the system SMC126-1-210 we could not detect any SMC component of the Na I D1 line because of strong blending with stellar lines.   

\begin{figure*}
\begin{minipage}[th]{0.5\linewidth}
\includegraphics[angle=0,scale=.51]{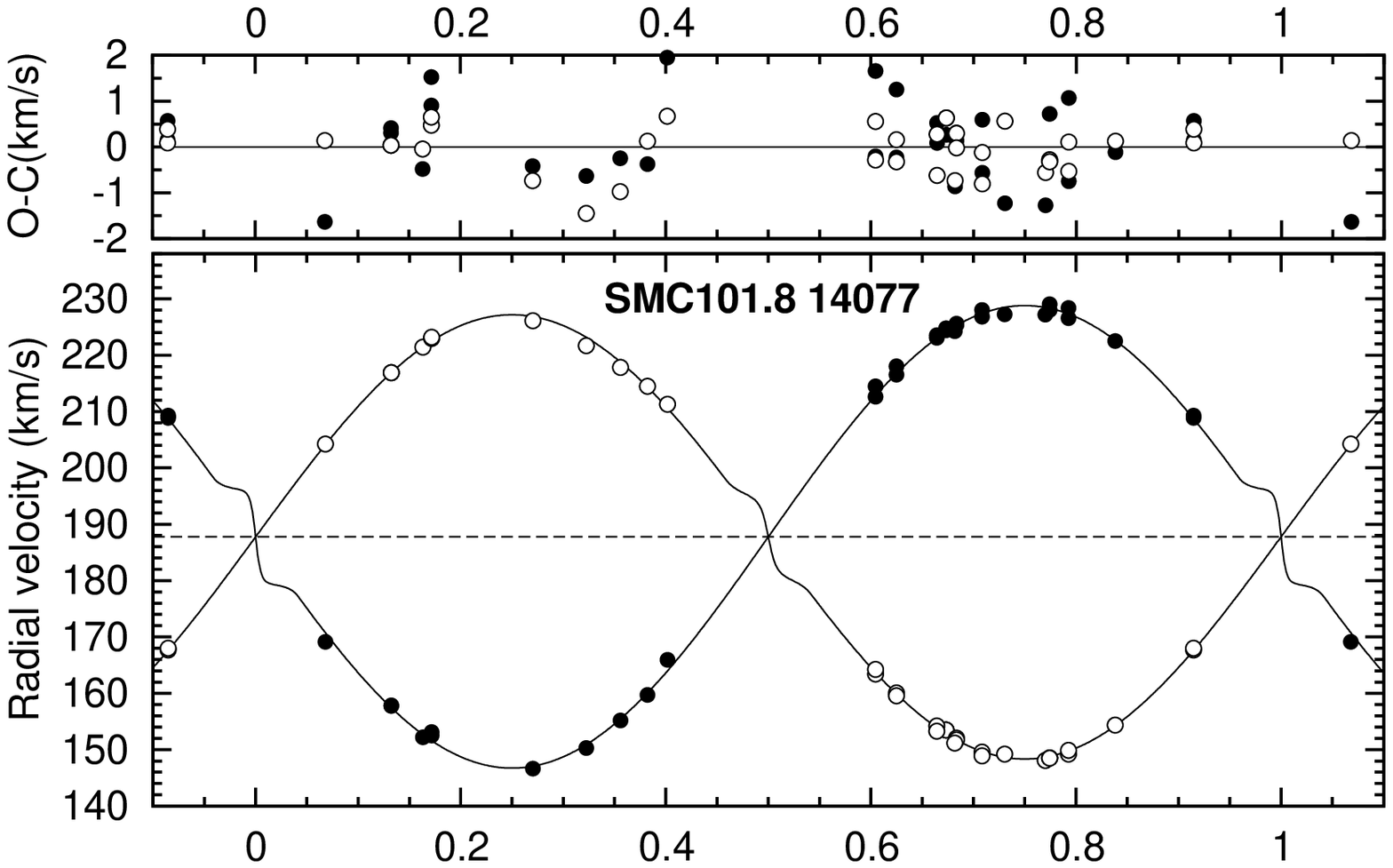} \vspace{-1.17cm}
\mbox{}\\ 
\includegraphics[angle=0,scale=.51]{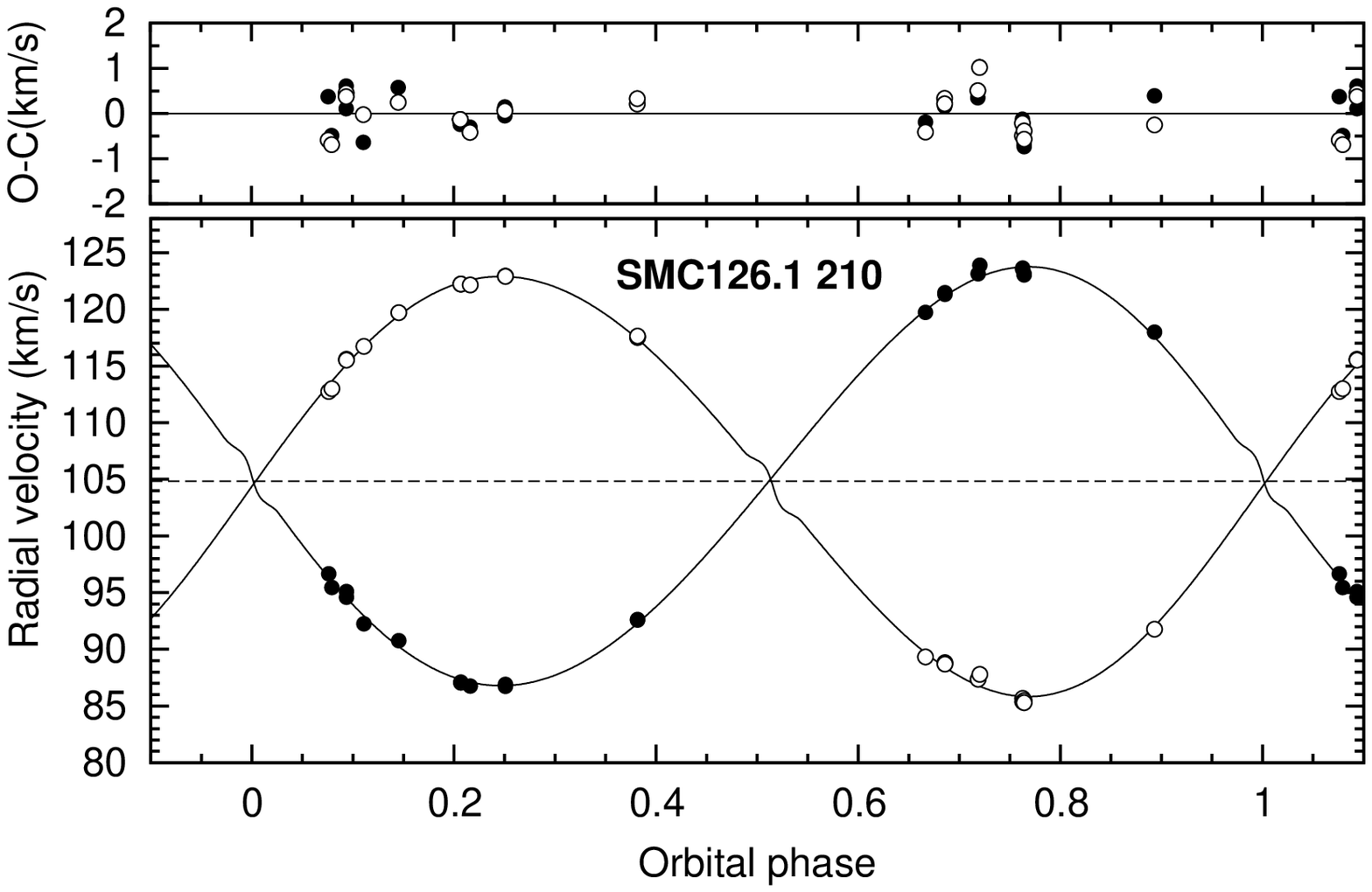}
\end{minipage}\hfill 
\begin{minipage}[th]{0.5\linewidth}
\includegraphics[angle=0,scale=.51]{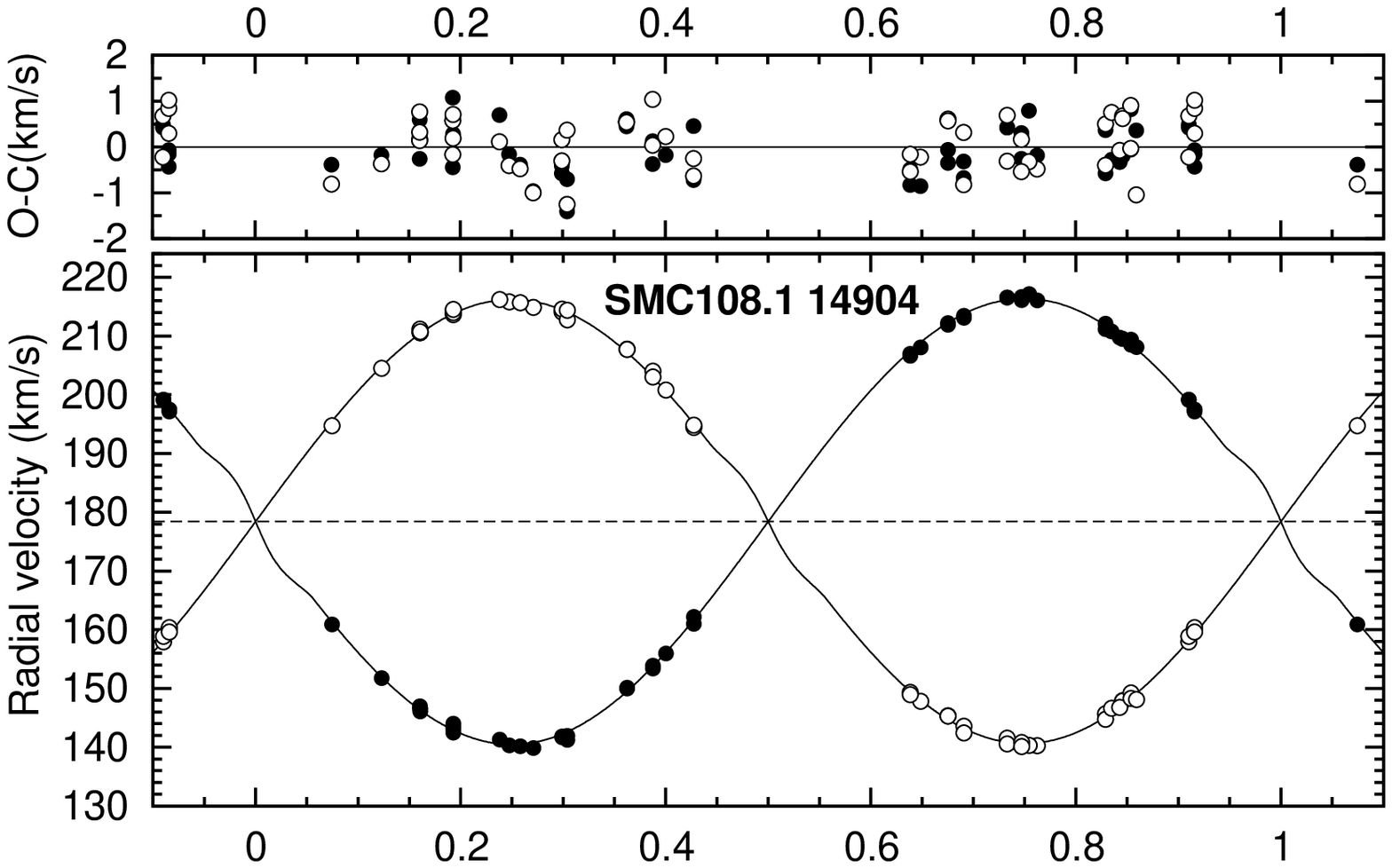} \vspace{-1.17cm}
\mbox{}\\ 
\includegraphics[angle=0,scale=.51]{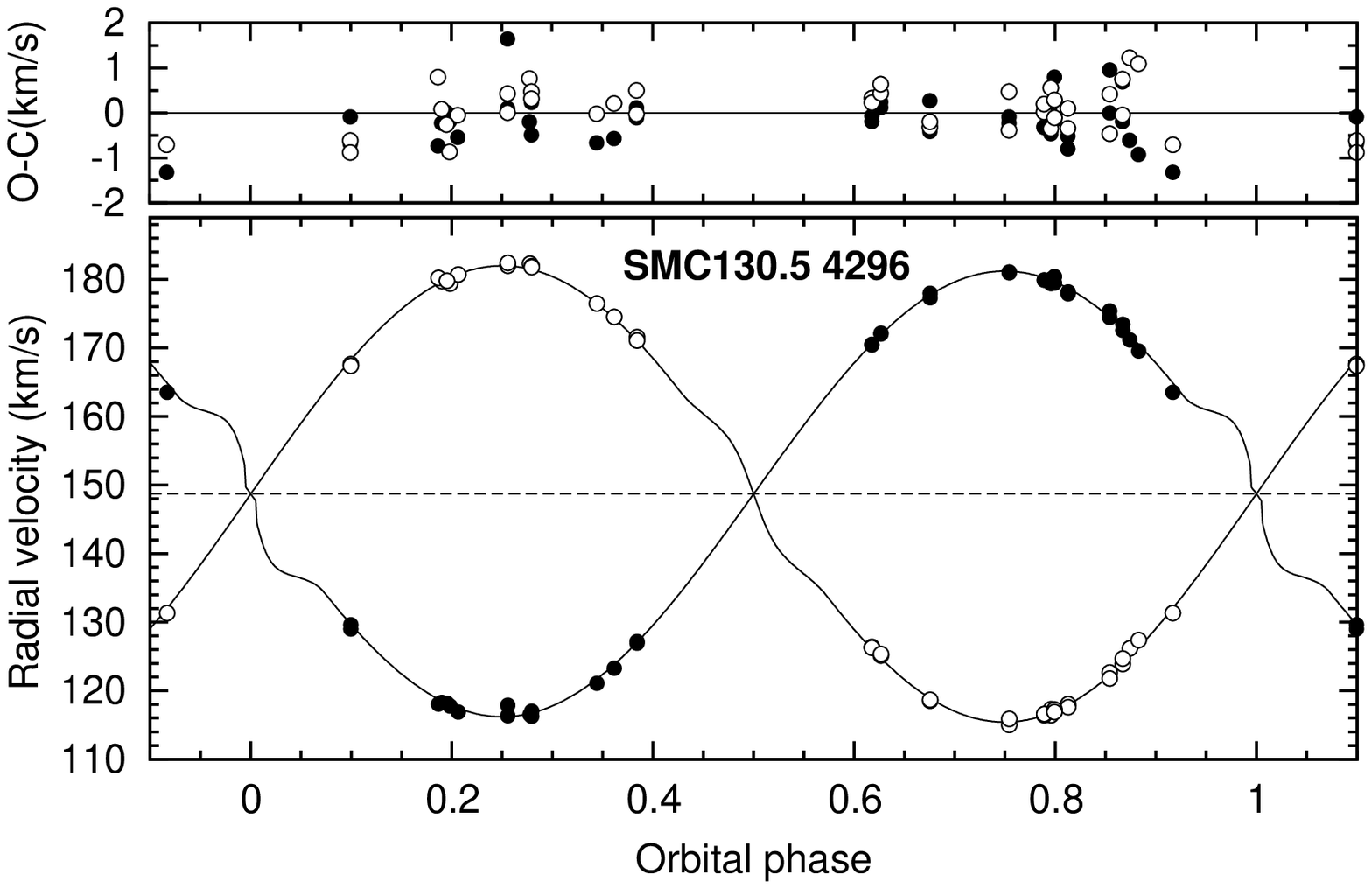}
\end{minipage}\hfill 
\caption{ The radial velocity solutions to four eclipsing binaries in the SMC. Filled circles correspond to the velocities of the primaries. The continuous line is the synthetic fit from the WD analysis.  Distortions near the photometric eclipses are caused by the Rossiter-McLaughlin effect. The dashed lines signify the systemic velocities.\label{fig3}}
\end{figure*}

The final method uses the effective temperatures we derived from the atmospheric analysis described in Section~\ref{atmo}. From the effective temperature - ($V\!-\!I$) color calibrations of Houdashelt et al. (2000) and Worthey \& Lee (2011) we estimated the intrinsic ($V\!-\!I$) colors of each component. These colors were compared with the observed colors of the components obtained from the preliminary solution to directly derive E($V\!-\!I$) color excesses. The reddening to each system was calculated as the mean value of the two components with the exception of the system SMC101.8 14077 where we have at our disposal a reddening estimate from only the secondary star.

Each of the three methods has an accuracy of approximately 0.03 - 0.04 mag. We calculated an average reddening for each system  from the three estimates, and used this new reddening estimate to update the temperature scale of the components. We then calculated new models and repeated the reddening derivation using the third method. The fourth column of Table~\ref{tbl-5} gives the reddening estimates after two such iterations. The fifth column presents the adopted E($B\!-\!V$) for each system as used in the "final" solution. We assigned a statistical error of 0.02 mag and an additional 0.02 mag systematic error for each estimate of the reddening.

\section{Adopted solutions}
\label{fin}

\begin{figure*}
\begin{minipage}[th]{0.5\linewidth}
\includegraphics[angle=0,scale=.51]{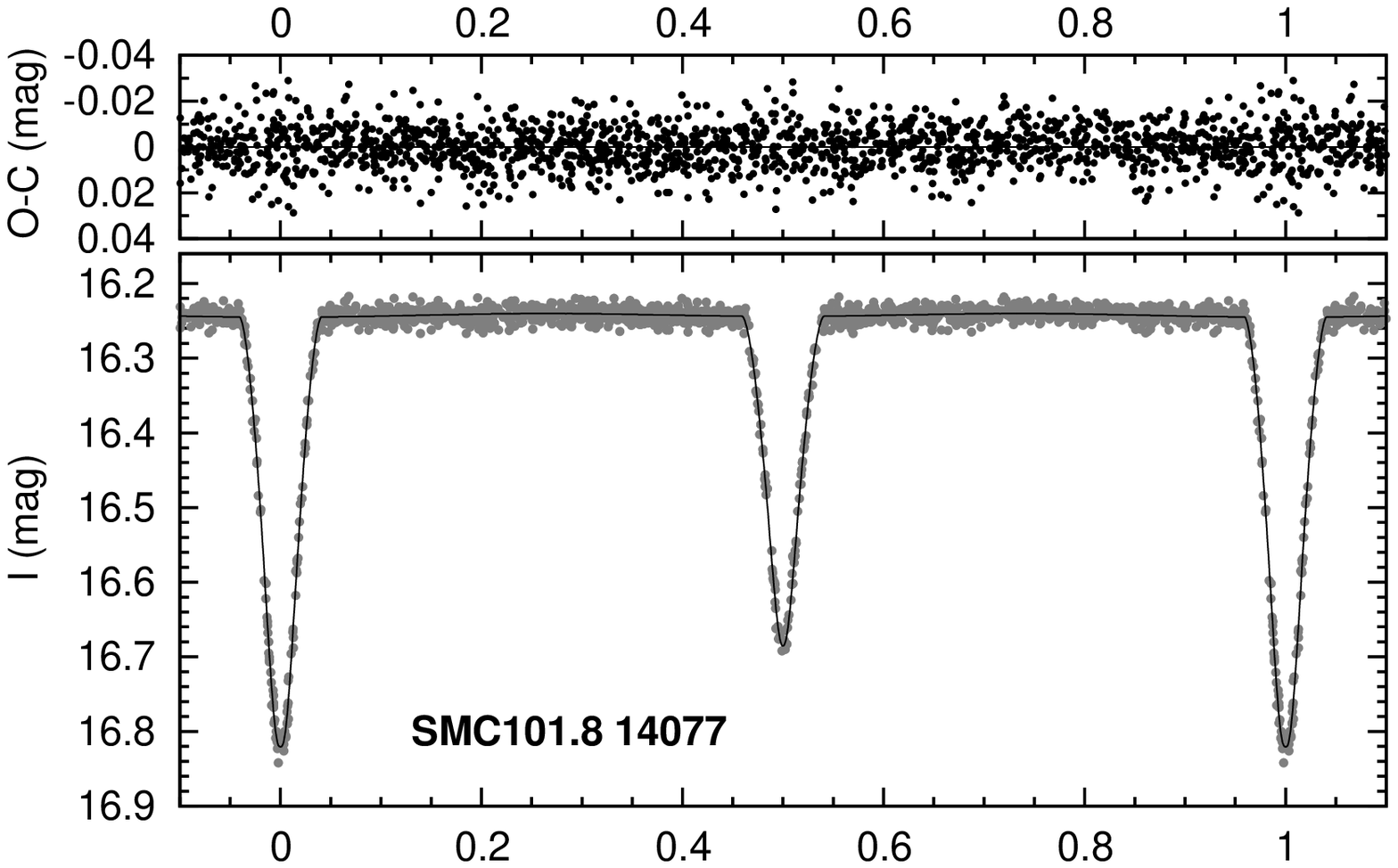} \vspace{-1.17cm}
\mbox{}\\ 
\includegraphics[angle=0,scale=.51]{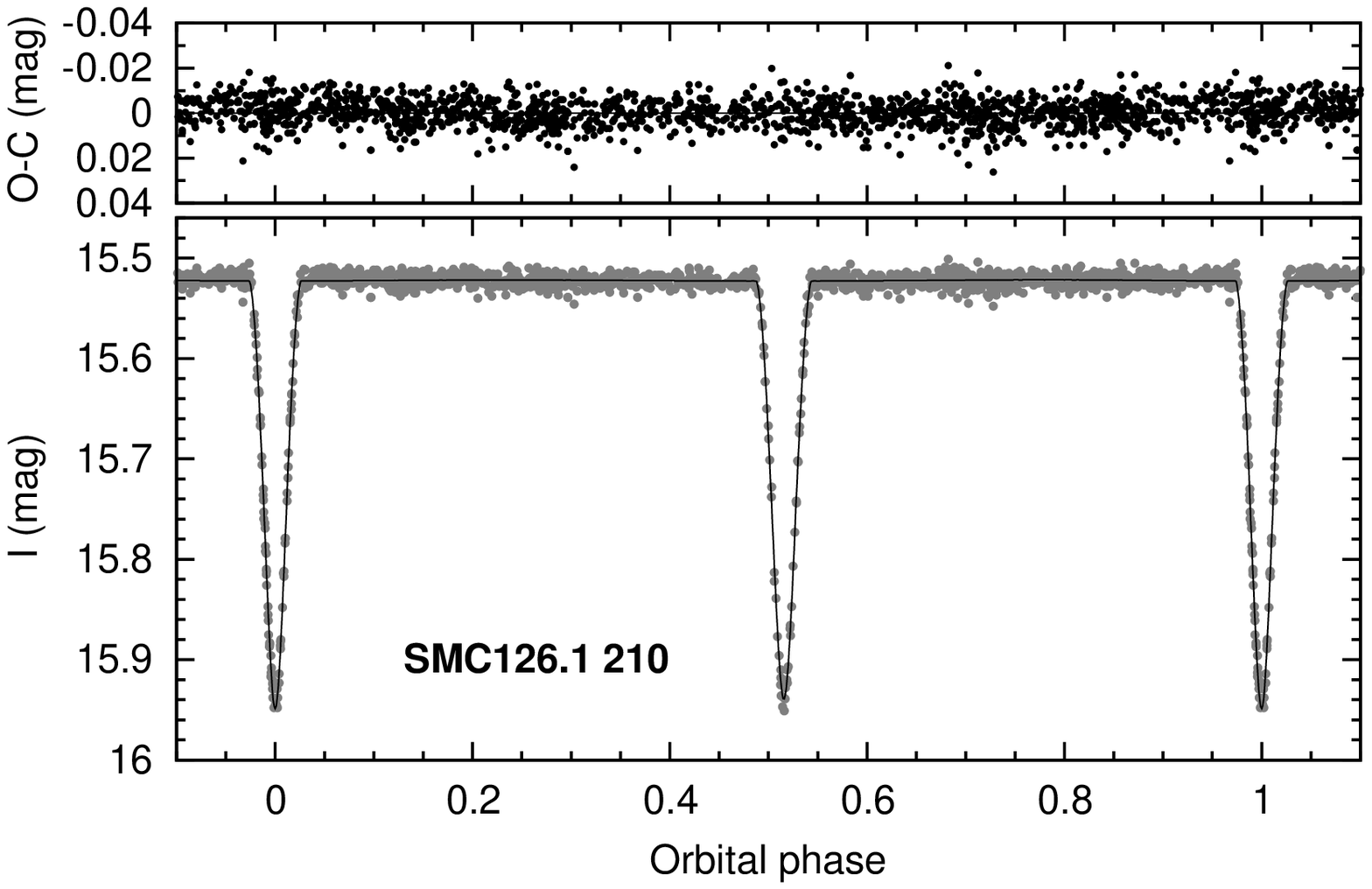}
\end{minipage}\hfill 
\begin{minipage}[th]{0.5\linewidth}
\includegraphics[angle=0,scale=.51]{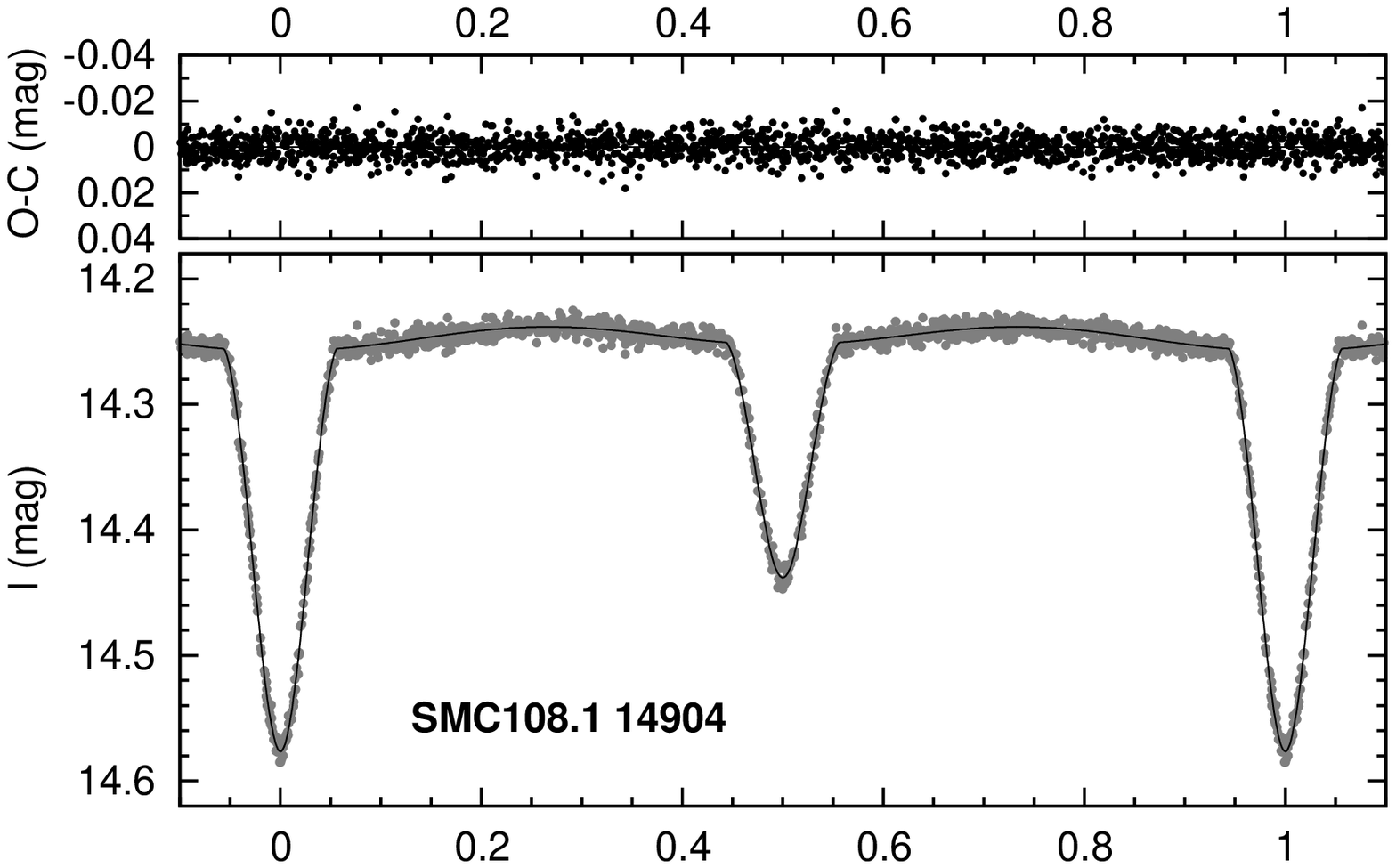} \vspace{-1.17cm}
\mbox{}\\ 
\includegraphics[angle=0,scale=.51]{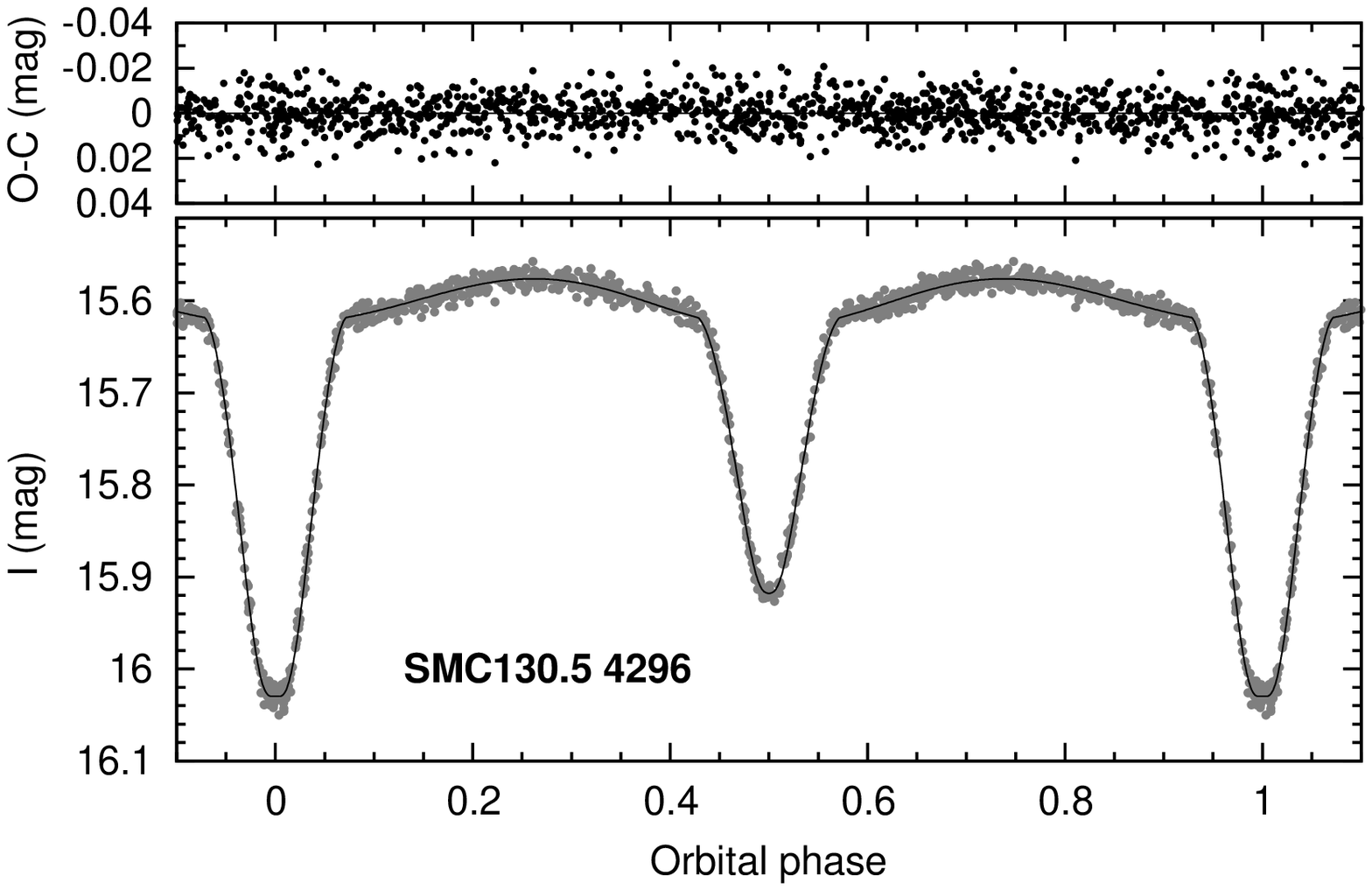}
\end{minipage}\hfill 
\caption{ The I-band light curve solutions to four eclipsing binaries in the SMC.\label{fig1}}
\end{figure*}

After the last reddening estimate we set the temperature of the primary in the way described in Section \ref{par} and we then recalculated all of the models. We call these the "basic" set of models. The temperature scale consistency of each eclipsing binary was checked by computing the distance to each system resulting from a scaling of the bolometric flux observed at Earth. To calculate the bolometric corrections we used an average from several calibrations \citep{cas10,mas06,alo99,flo96}. These distances were then compared to the distances computed with a surface brightness - color relation. In all cases both estimates of the  distances agree within the errors, thus confirming the proper temperature scales.   

Finally we computed three additional sets of models by adjusting: 1) the four coefficients of the linear law of limb darkening (LB=+1) for both stars and both light curves, 2)  the third light $l_3$ in two bands, and 3) the third light and linear coefficients of limb darkening together - six more free parameters. We compared the model  basic set with these new sets to find the physical model with the lowest reduced $\chi^2_{\rm r}$. Models with unphysical values of the third light ($l_3\!<\!0$) and/or with very high limb darkening coefficients ($x\!>\!1.1$) were excluded even if they produced a better formal fit with lower $\chi^2_{\rm r}$.  Table~\ref{tbl-3} lists the parameters of the best model for each system. The quoted uncertainties are errors calculated with the DC routine. The quantity $P_{\rm orb}$ signifies the rest frame orbital period. 

\begin{figure*}
\begin{minipage}[th]{0.5\linewidth}
\includegraphics[angle=0,scale=.51]{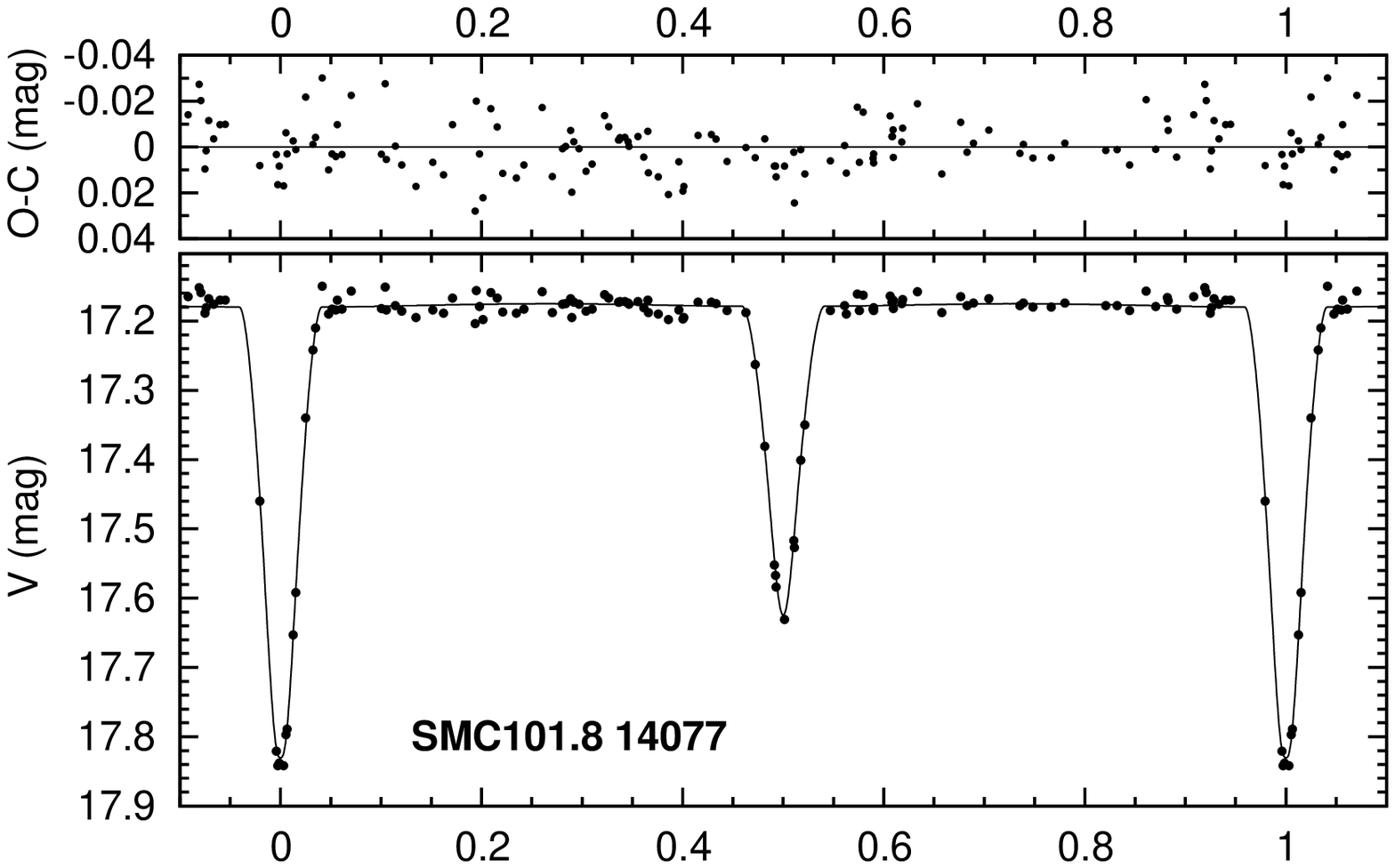} \vspace{-1.17cm}
\mbox{}\\ 
\includegraphics[angle=0,scale=.51]{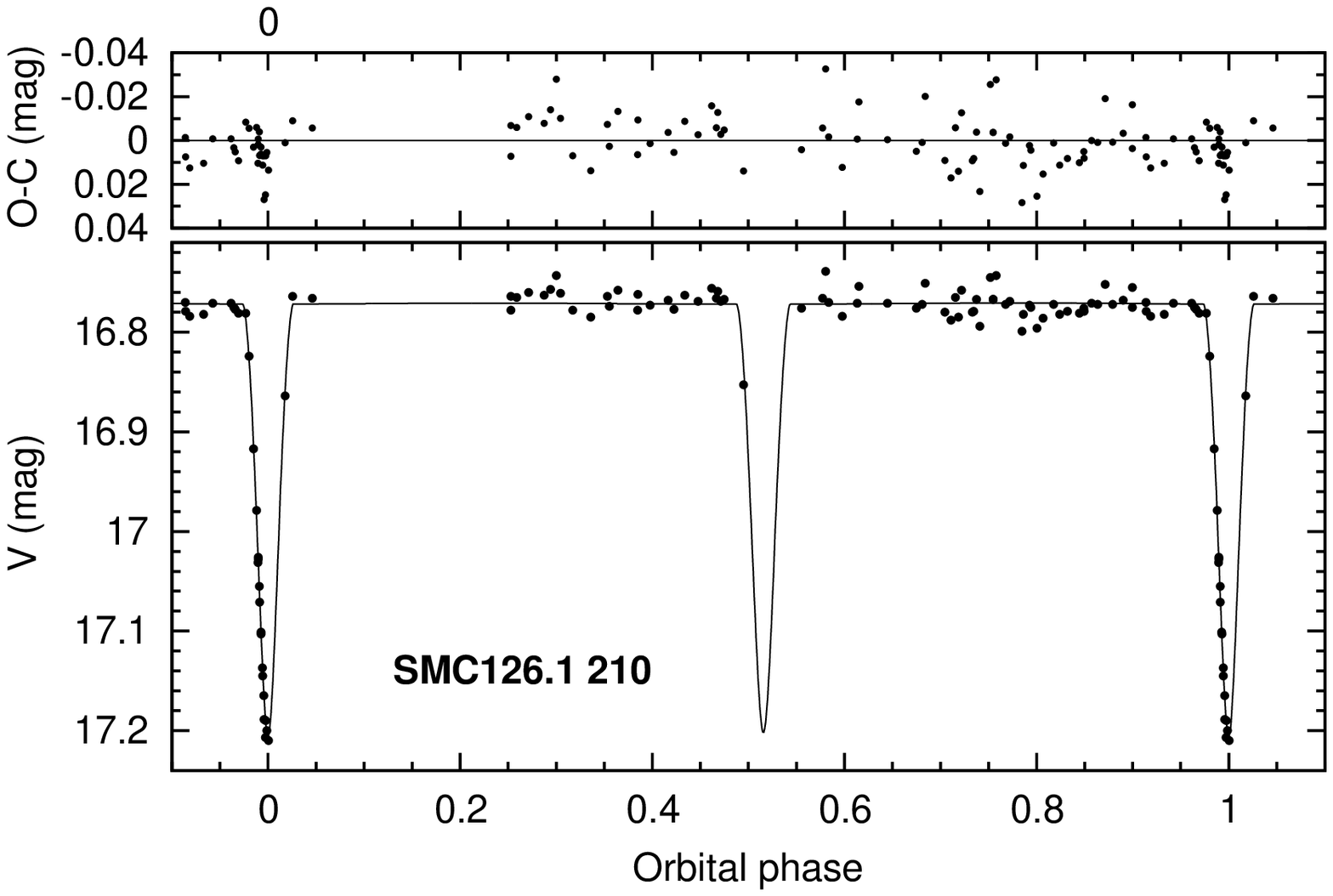}
\end{minipage}\hfill 
\begin{minipage}[th]{0.5\linewidth}
\includegraphics[angle=0,scale=.51]{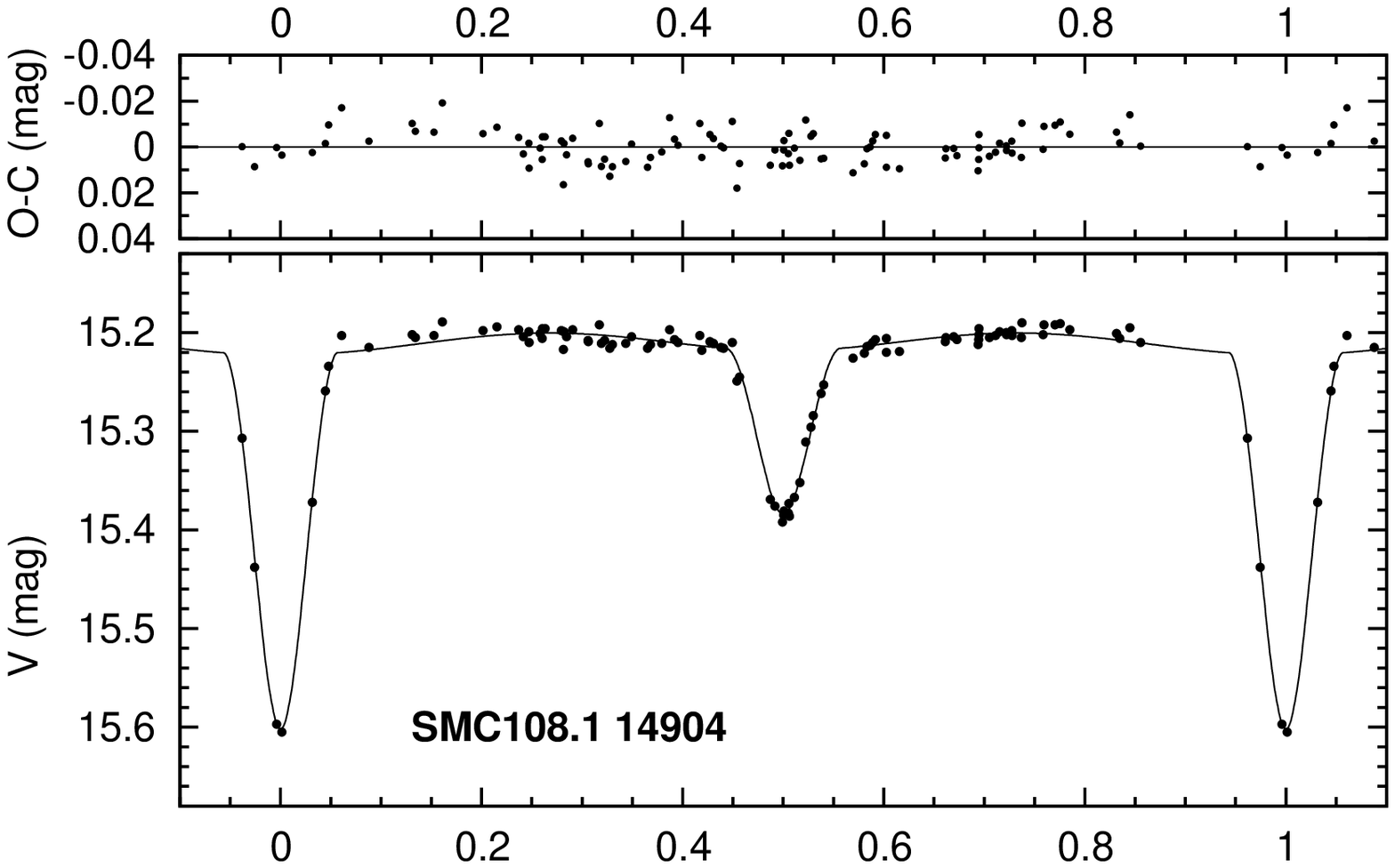} \vspace{-1.17cm}
\mbox{}\\ 
\includegraphics[angle=0,scale=.51]{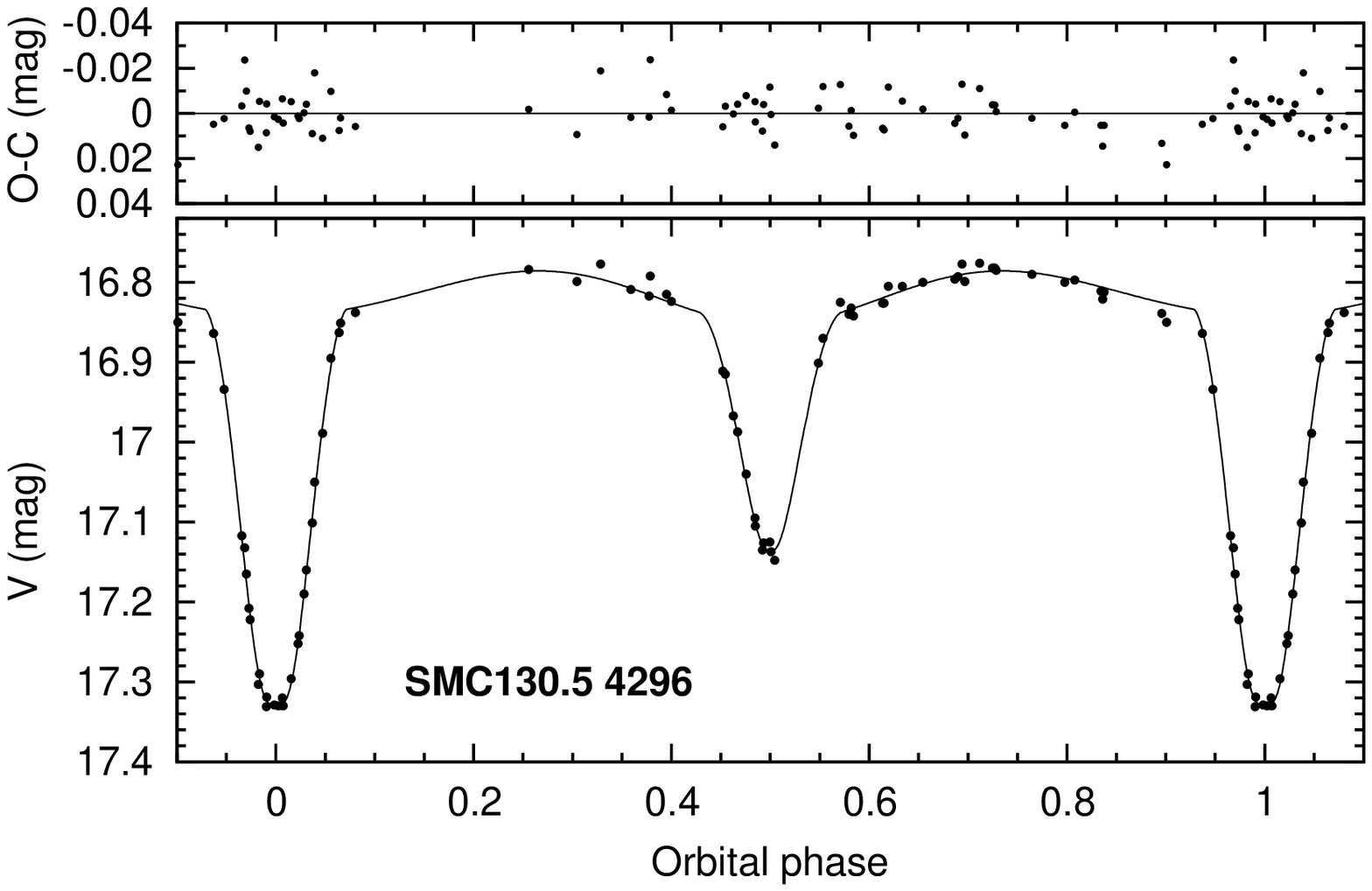}
\end{minipage}\hfill 
\caption{ The V-band light curve solutions to four eclipsing binaries in the SMC.\label{fig2}}
\end{figure*}

\begin{figure*}
\begin{minipage}[th]{0.5\linewidth}
\includegraphics[angle=0,scale=.5]{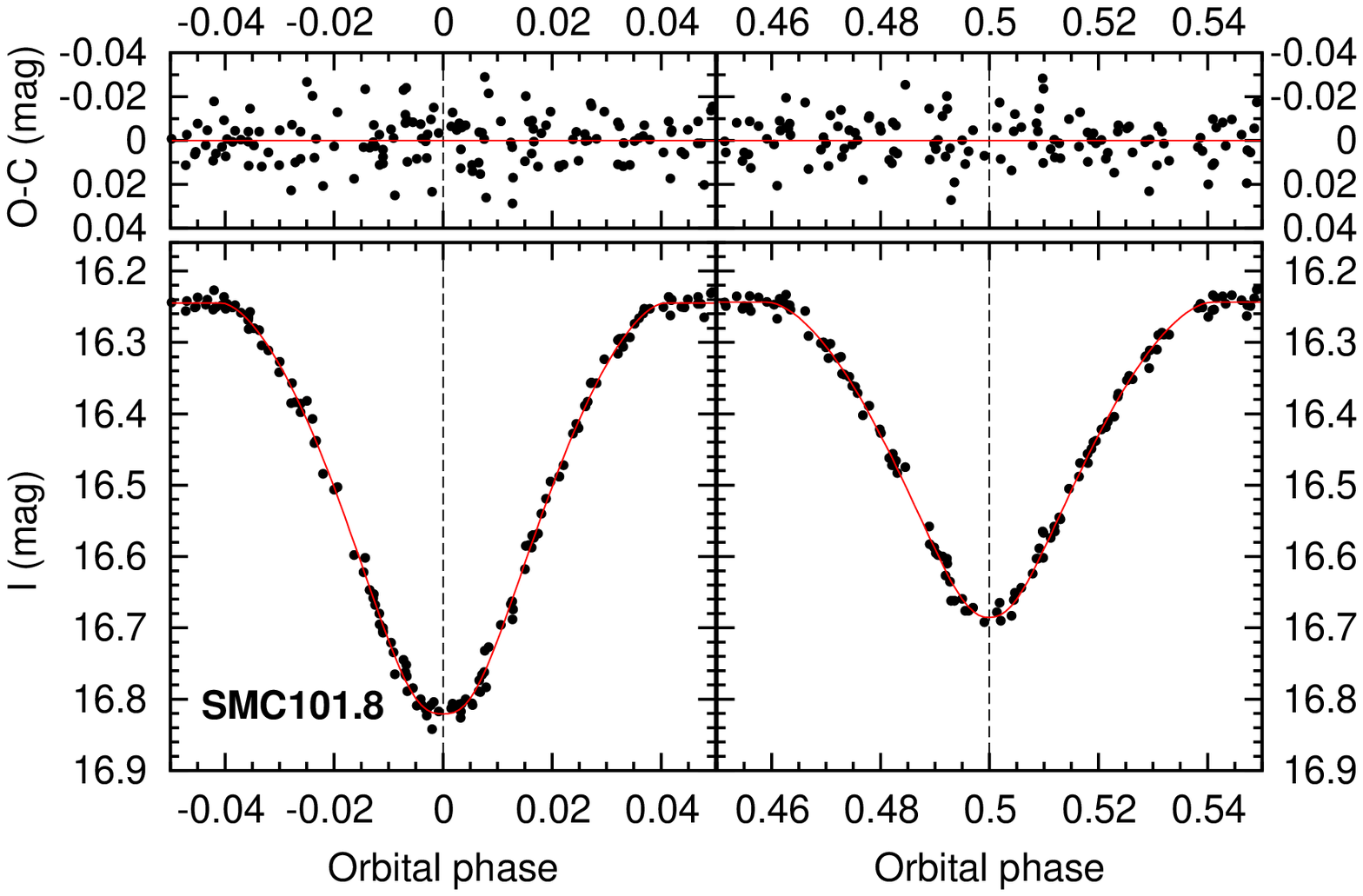} \vspace{-.17cm}
\mbox{}\\ 
\includegraphics[angle=0,scale=.5]{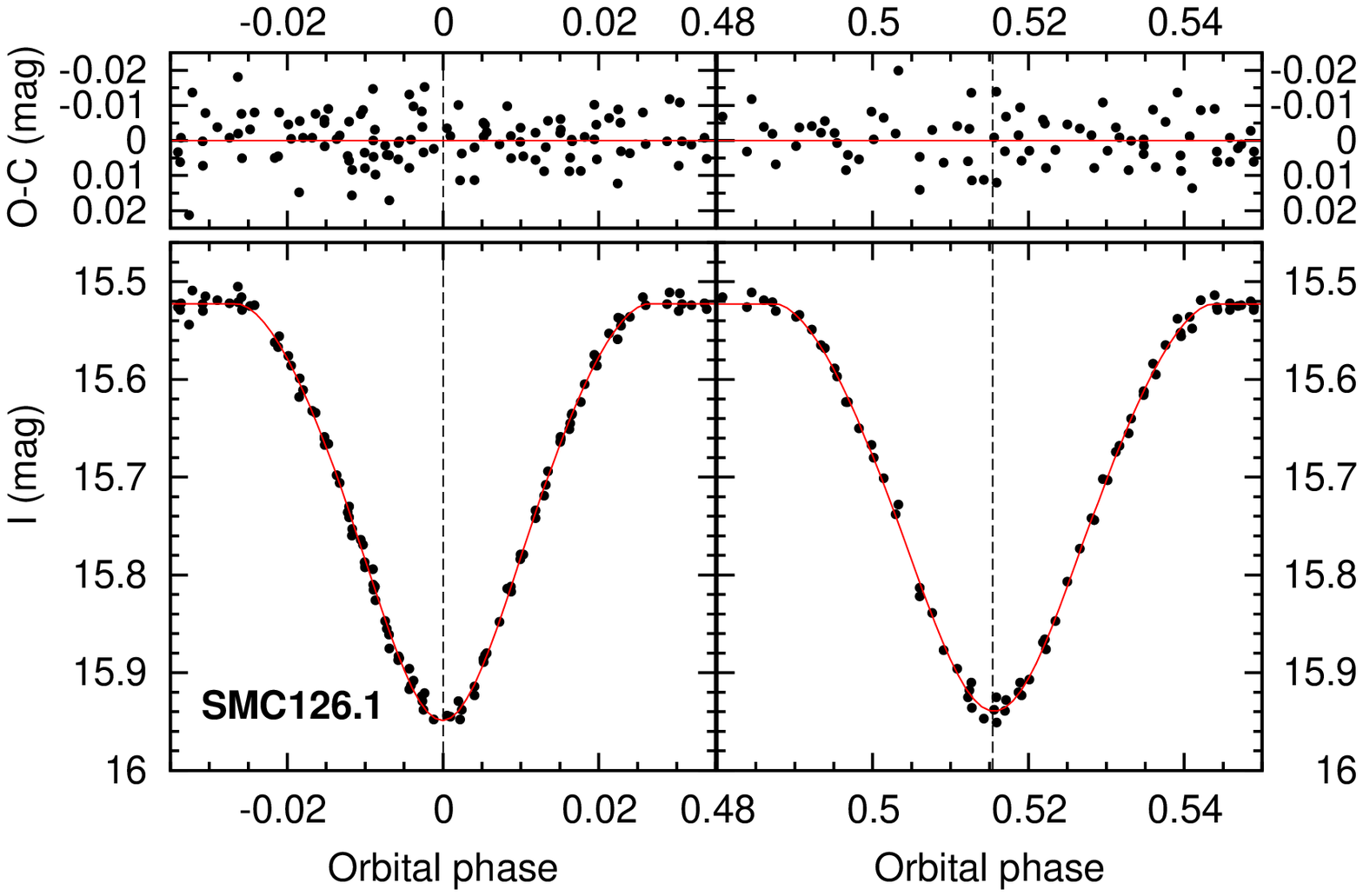}
\end{minipage}\hfill 
\begin{minipage}[th]{0.5\linewidth}
\includegraphics[angle=0,scale=.5]{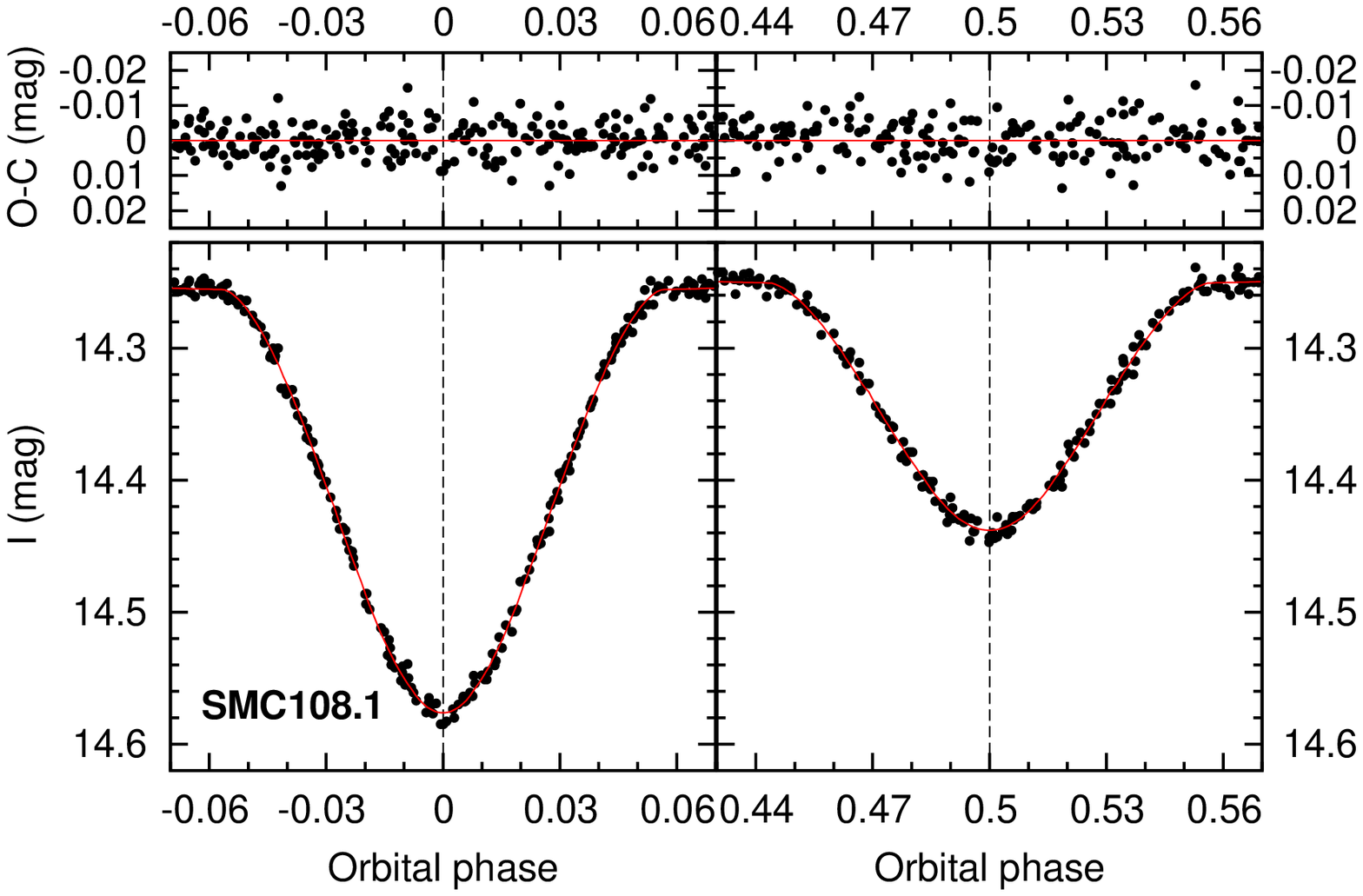} \vspace{-.17cm}
\mbox{}\\ 
\includegraphics[angle=0,scale=.5]{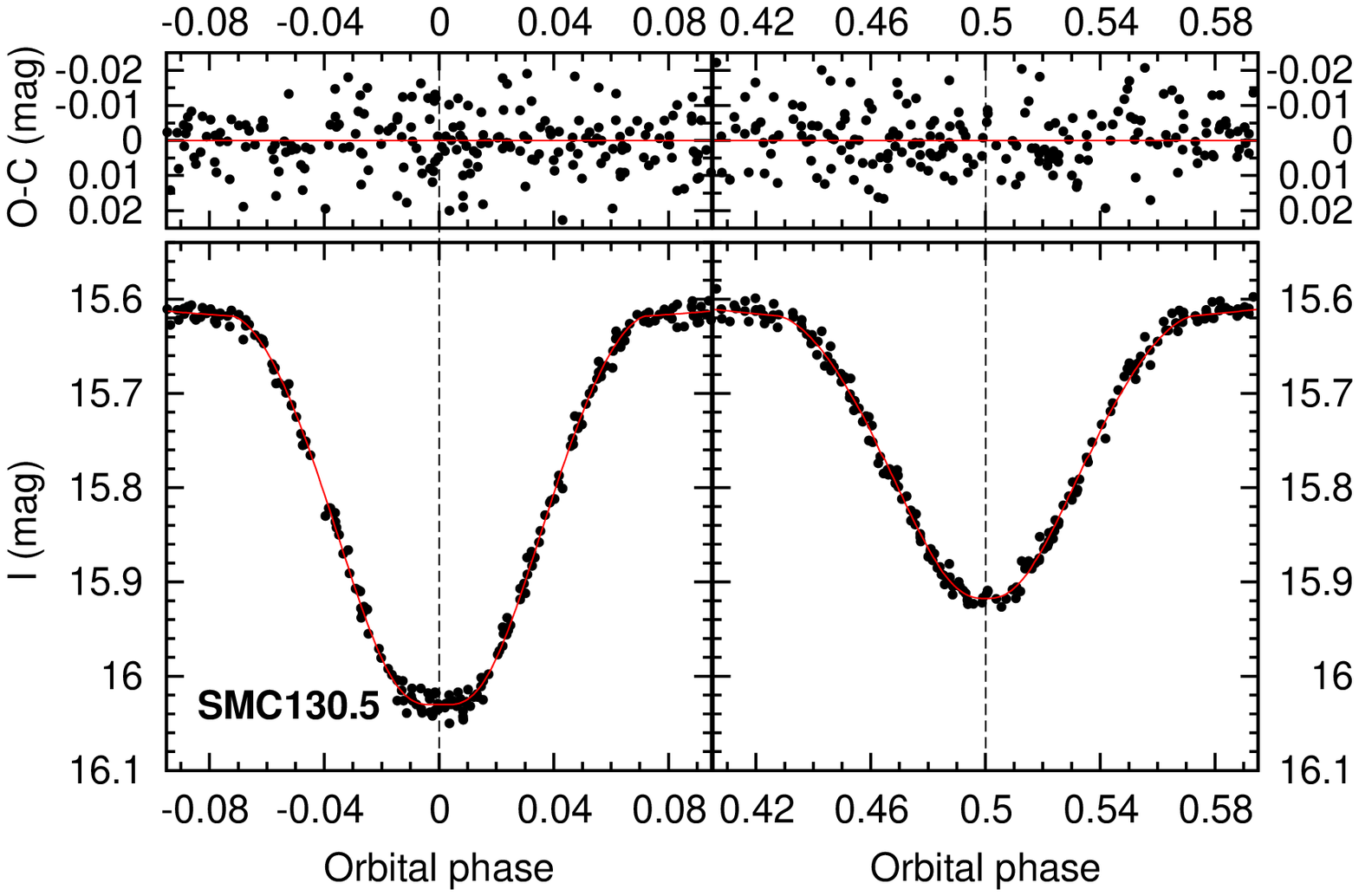}
\end{minipage}\hfill 
\caption{ The close-up of the eclipses of individual eclipsing binaries in the SMC. Continuous red line is a best-fit model from the WD program, vertical dashed lines mark times of mid-eclipse.\label{fig6}}
\end{figure*}

Solutions to the radial velocity curves  are presented in Fig.~\ref{fig3}.  The I-band and V-band light curve solutions are presented in Figures~\ref{fig1} and~\ref{fig2}, respectively. Figure~\ref{fig6} shows some details of the shapes of the eclipses.    The absolute dimensions are reported in Table~\ref{tbl-6}. These were calculated following \cite{gra12} adopting the same values for physical constants. The spectral type of each component was estimated according to its effective temperature using a calibration by \cite{alo99}. We now comment on individual systems and particular model solutions. 

\begin{deluxetable*}{@{}lcccc}
\tabletypesize{\scriptsize}
\tablecaption{Model parameters from the Wilson-Devinney code\tablenotemark{a} \label{tbl-3}}
\tablewidth{0pt}
\tablehead{
\colhead{Parameter} & \colhead{SMC101.8 14077} &\colhead{SMC108.1 14904}&\colhead{SMC126.1 210}&\colhead{SMC130.5 4296} 
}
\startdata
 {\bf Orbital param.} &&&&\\
 $P_{\rm obs}$ (days) &$102.8984 \pm 0.0003$  & $185.2176 \pm 0.0017$ &$634.999 \pm 0.009$ &$120.4681 \pm 0.0009$ \\
 $T_0$ (HJD-2450000) &$2818.913 \pm 0.006$&$641.081 \pm 0.010$&$2858.30 \pm 0.04$&$2052.976 \pm 0.009$ \\ $a\sin{i}$ (R$_\odot$) &$163.69 \pm 0.47$ &$277.61 \pm 0.45$ &$464.35 \pm 1.90$&$157.10 \pm 0.40$ \\
 $q=M_2/M_1$ &$1.0404 \pm 0.0062$ &$1.0028 \pm 0.0030$ &$0.9971 \pm 0.0078$ &$0.9736 \pm 0.0048$ \\
 $\gamma$ (km s$^{-1}$) &$187.76 \pm 0.11$ &$178.42 \pm 0.07$&$104.82 \pm 0.08$&$148.70 \pm 0.09$\\
 $e$ &   $0$ & $0$ &$0.0422 \pm 0.0018$ & 0\\
 $\omega$ (deg)& $90$  & $90$ &$54.4 \pm 1.7$& 90\\
 &&&&\\
 {\bf Photometric param.} &&&&\\
 $i$ (deg) & $88.04 \pm 0.23$ & $78.87 \pm 0.10$ &$86.92 \pm 0.09$ &$83.09 \pm 0.13$ \\
 $T_2/T_1$ &$0.9269 \pm 0.0014$ &$0.8701 \pm 0.0054$&$1.0065 \pm 0.0009$ &$0.9194 \pm 0.0055$ \\
 $r_1$ & $0.1093 \pm 0.0012$ & $0.1660 \pm 0.0016$ &$0.0936 \pm 0.0011 $&$0.1608 \pm 0.0012$\\
 $r_2$ & $ 0.1458 \pm 0.0012$ & $0.2265 \pm 0.0014$&$0.0839 \pm 0.0016 $&$0.2908 \pm 0.0015$\\
 &&&&\\
 $(L2/L1)_V$ & $1.205 \pm 0.010$ & $0.928 \pm 0.013$&$0.838 \pm 0.016$&$1.939 \pm 0.015$ \\
 $(L2/L1)_I$ & $1.349 \pm  0.008$ &$ 1.169 \pm 0.013$ &$0.827 \pm 0.016$&$2.370 \pm 0.009$\\
 $(L2/L1)_K$ & $1.620$\tablenotemark{b} &$1.572$\tablenotemark{d} &0.810\tablenotemark{b}&2.925\tablenotemark{d}\\
 $x_{1,V}$  &- &$0.873 \pm 0.116$&-&$0.764 \pm 0.100$ \\
 $x_{2,V}$  &- &$0.874 \pm 0.106$&-&$0.873 \pm 0.048$\\
 $x_{1,I}$ &-&$0.461 \pm 0.050$&-&$0.597 \pm 0.070$ \\
 $x_{2,I}$ &-&$0.425 \pm 0.045$&-&$0.562 \pm 0.030$ \\
 $l_{3,V}$  & $0.005 \pm 0.018$ & 0&0& 0\\
 $l_{3,I}$ & $0.043 \pm 0.017$&0&0& 0\\
 $l_{3,K}$ &0.043\tablenotemark{c} &0&0&0\\
 &&&&\\
 {\bf Derived quantities} &&&&\\
 $P_{\rm orb}$ (days) &$102.8340 \pm 0.0003$&$185.1076 \pm 0.0017$&$634.777 \pm 0.009$ &$120.4084 \pm 0.0009$\\
 $a$ (R$_\odot$) &$163.69 \pm 0.47$&$282.76 \pm 0.45$&$464.86 \pm 1.90$&$158.17 \pm 0.40$ \\
 $K_1$ (km s$^{-1}$) &$41.03 \pm 0.20$&$37.96 \pm 0.08$&$18.48 \pm 0.11$&$32.54 \pm 0.12$\\
 $K_2$ (km s$^{-1}$) &$39.44 \pm 0.12$&$37.85 \pm 0.09$&$18.54 \pm 0.10$&$33.42 \pm 0.11$\\
 rms$_{1}$ (km s$^{-1}$) &0.84& 0.52& 0.44& 0.56 \\ 
 rms$_{2}$ (km s$^{-1}$) &0.52 & 0.57& 0.42& 0.51 \\
 &&&&\\
 $(r_1+r_2)$ &$ 0.2551$& $0.3926 $&0.1775&0.4516\\
 $k=r_2/r_1$& $1.3332$& $ 1.3642 $&0.8961&1.808\\
 $(j_2/j_1)_V$  &0.678&0.499&1.044&0.593\\
 $(j_2/j_1)_I$ &0.759&0.628&1.029&0.725
 \enddata
\tablenotetext{a}{Simultaneous solution of V-band and I-band light curves together with radial velocity curves of both components. Uncertainties quoted are the standard errors from Differential Corrections subroutine.}
\tablenotetext{b}{Extrapolated from the WD model}
\tablenotetext{c}{Assumed}
\tablenotetext{d}{The resulting light ratio assuming the same distance to both stars}
\end{deluxetable*}

\begin{deluxetable*}{@{\hspace{-12pt}}lcccccccc}
\tabletypesize{\scriptsize}
\tablecaption{Physical Properties of the SMC Eclipsing Binaries \label{tbl-6}}
\tablewidth{0pt}
\tablehead{
\colhead{Property\tablenotemark{a,b}} &\multicolumn{2}{c}{{SMC101.8 14077}}& \multicolumn{2}{c}{{SMC108.1 14904}}& \multicolumn{2}{c}{{SMC126.1 210}}& \multicolumn{2}{c}{{SMC130.5 4296}} \\
&\colhead{Primary} & \colhead{Secondary} &\colhead{Primary} & \colhead{Secondary} &\colhead{Primary} & \colhead{Secondary} &\colhead{Primary} &\colhead{Secondary}
}
\startdata
 Spectral Type  &  G0 III & G4 III &F9 II& G7 II&K2 III&K1 III&G7 III & K1 III\\
 $V$ (mag) & $18.041\pm0.013$  & $17.838\pm0.013$  &$15.918\pm0.010$&$15.999\pm0.010$&$17.432\pm0.011$&$17.624\pm0.011$&$17.954\pm0.014$&$17.235\pm0.012$\\
 \mbox{$V\!-\!I$} (mag) & $0.825\pm0.016$ & $0.947\pm0.016$ &$0.835\pm0.014$&$1.086\pm0.014$&$1.256\pm0.015$&$1.241\pm0.015$&$1.058\pm0.017$&$1.276\pm0.016$\\
 \mbox{$V\!-\!K$} (mag) & $1.828\pm0.020$ & $2.149\pm0.020$ &$1.839\pm0.018$&$2.411\pm0.018$&$2.916\pm0.017$&$2.879\pm0.017$&$2.432\pm0.022$&$2.878\pm0.020$\\
 \mbox{$J\!-\!K$} (mag) & $0.481\pm0.025$ & $0.575\pm0.025$ &$0.536\pm0.021$&$0.662\pm0.021$&$0.846\pm0.024$&$0.836\pm0.024$&$0.765\pm0.027$&$0.852\pm0.025$\\
Mass (M$_\sun$) & $2.725\pm 0.034$ & $2.835 \pm 0.055$ &$4.416 \pm 0.041$&$4.429 \pm 0.037$&$1.674 \pm 0.037$&$1.669 \pm 0.039$ &$1.854 \pm 0.025$&$1.805 \pm 0.027$\\
 Radius (R$_\sun$) & $17.90 \pm 0.50$& $23.86 \pm 0.31$&$46.95 \pm 0.53$&$64.05 \pm 0.50$&$43.52 \pm 1.02$ &$39.00 \pm 0.98$&$25.44 \pm 0.25$&$46.00 \pm 0.35$\\ 
 $\log g$ (cgs) & $2.368 \pm 0.029$ & $2.136 \pm 0.019$ &$1.740 \pm 0.014$&$1.472 \pm 0.010$ &$1.385 \pm 0.029$ &$1.479 \pm 0.031$ &$1.895 \pm 0.014$&$1.369 \pm 0.013$\\
 $T_{\rm eff}$ (K) & $5580 \pm 95$ & $5170 \pm 90$ &$5675 \pm 105$ &$4955 \pm 90$&$4480 \pm 70$ &$4510 \pm 70$ &$4912 \pm 80 $& $4515 \pm 75$\\
 Luminosity (L$_\sun$) & $280 \pm 34$& $365 \pm  35$ &$2055 \pm 200$ &$2220 \pm 195$ &$685 \pm 74$ &$565 \pm 63$ &$338 \pm 28$ &$790 \pm 65$\\
 $M_{\rm bol}$ (mag) &$-1.38$ & $-1.67$ &$-3.54$&$-3.63$&$-2.35$&$-2.14$&$-1.57$&$-2.50$\\
 $M_V$ (mag) &$-1.24$ & $-1.44$   &$-3.42$&$-3.34$&$-1.81$&$-1.62$&$-1.26$&$-1.98$\\
 $\left[{\rm Fe/H}\right]$ (dex)& - &$-1.01$&$-0.95$&$-0.64$&$-0.94$&$-0.79$&$-0.77$&$-0.99$ \\
 E(\bv) (mag)& \multicolumn{2}{c}{$0.067 \pm 0.020$}&\multicolumn{2}{c}{$0.093 \pm 0.020$}&\multicolumn{2}{c}{$0.080 \pm 0.020$}&\multicolumn{2}{c}{$0.079 \pm 0.020$}
 \enddata
 \tablenotetext{a}{Absolute dimensions were calculated assuming: $G=6.673\cdot10^{-8}$ cm$^3$g$^{-1}$s$^{-2}$, $R_\sun=695 600$ km, $M_\sun=1.9888\cdot10^{33}$ g, $T_{\rm eff,\sun}=5777$ K, $M_{bol,\sun}=+4.75$.}
\tablenotetext{b}{The magnitudes and colors are observed values.}
\end{deluxetable*}

\begin{figure}
\includegraphics[angle=0,scale=.50]{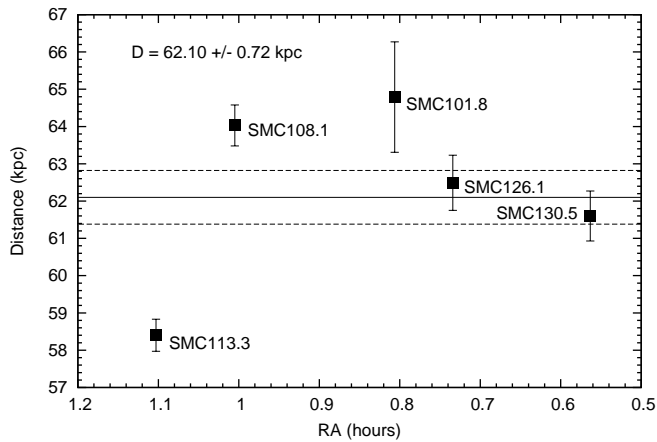}
\caption{ The distance estimates to four new eclipsing binaries in the SMC and the system SMC113.4 4007 published previously by our team. The continuous line is an unweighted mean distance from all five binaries and the dashed lines 
represent a 1-$\sigma$ range. The errorbars signify statistical uncertainties. \label{fig8}}
\end{figure}

\begin{figure*}
\begin{minipage}[th]{0.5\linewidth}
\includegraphics[angle=0,scale=.42]{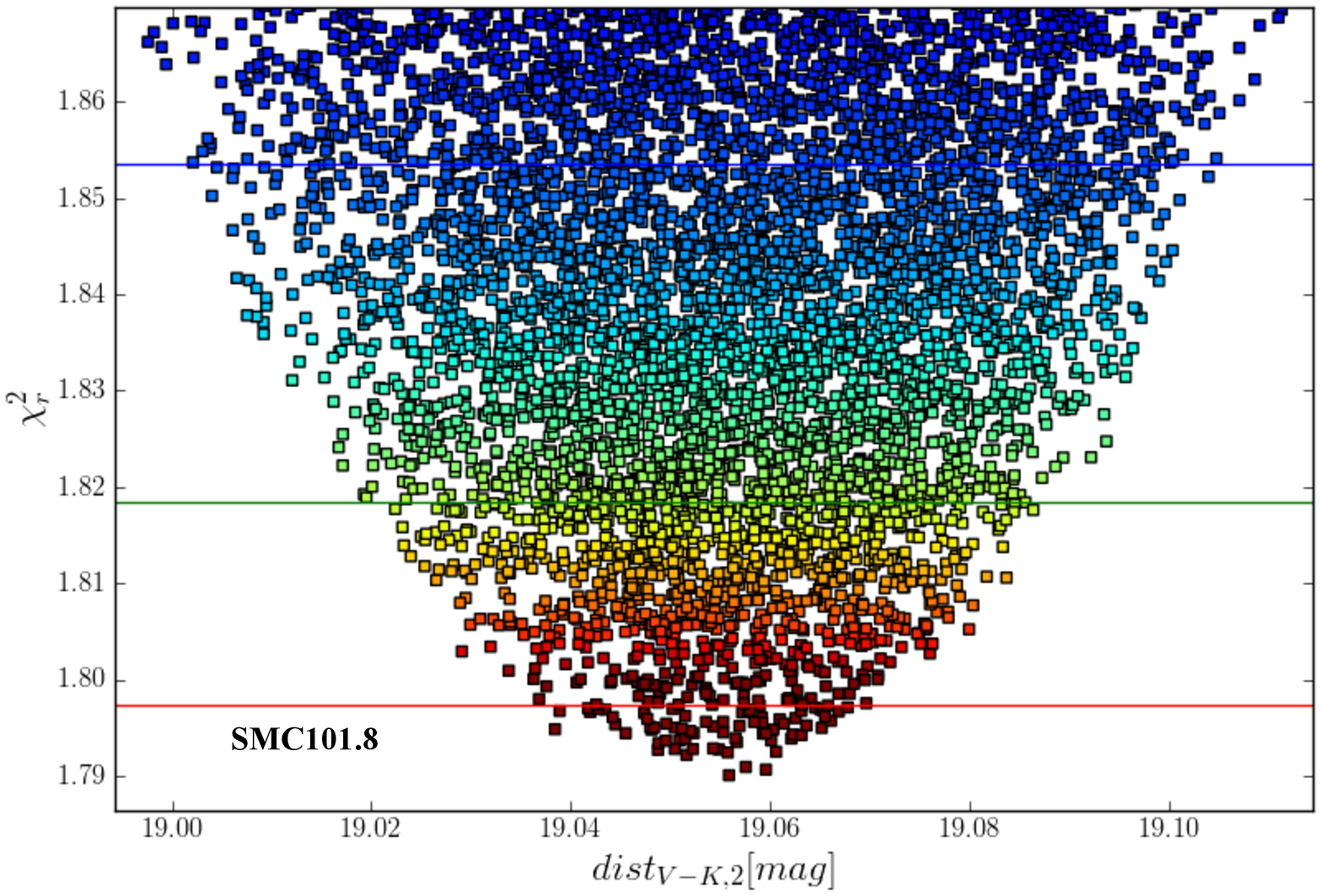}
\mbox{}\\
\includegraphics[angle=0,scale=.42]{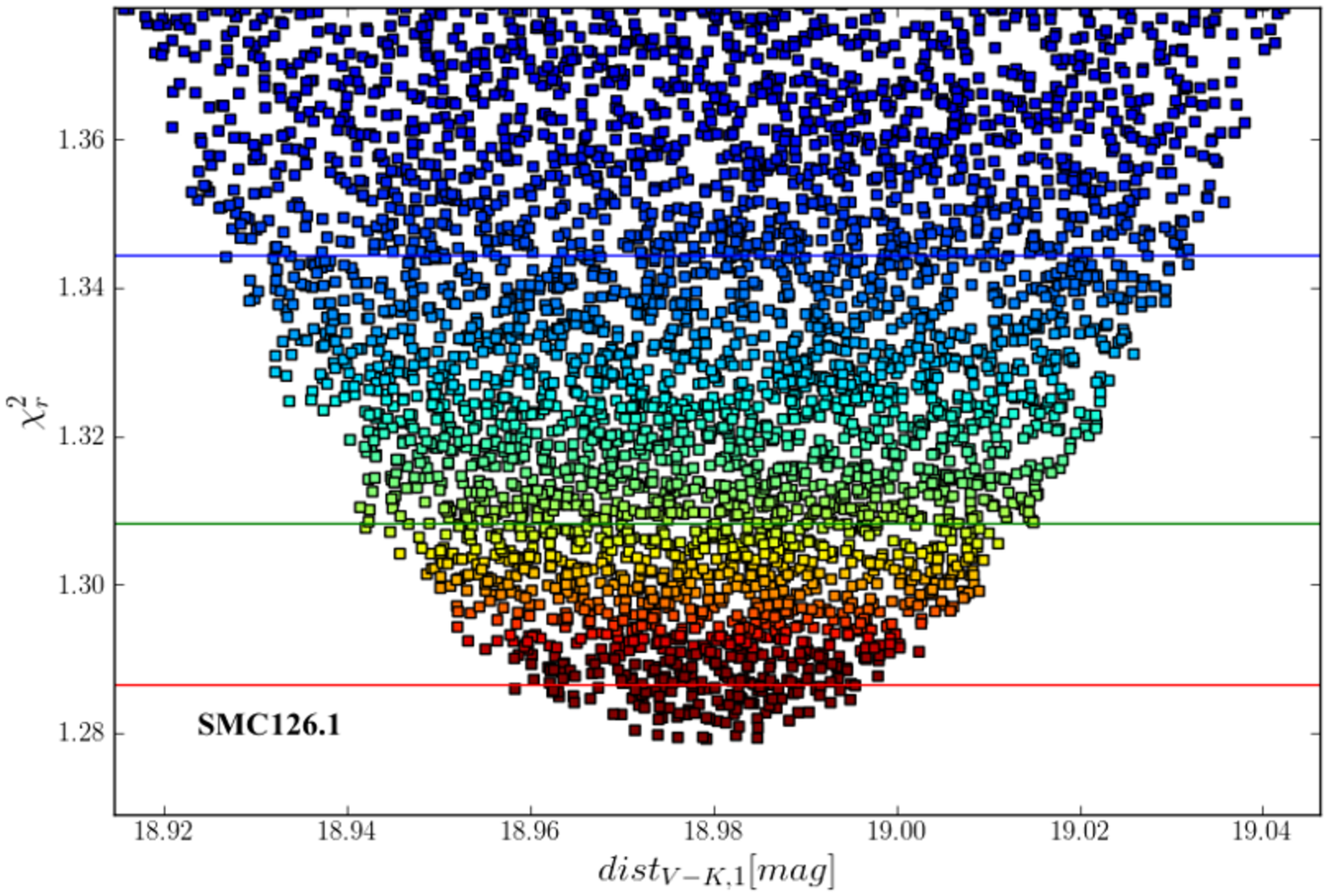}
\mbox{}\\
\end{minipage}\hfill 
\begin{minipage}[th]{0.5\linewidth} 
\vspace{-0.25cm}
\includegraphics[angle=0,scale=.42]{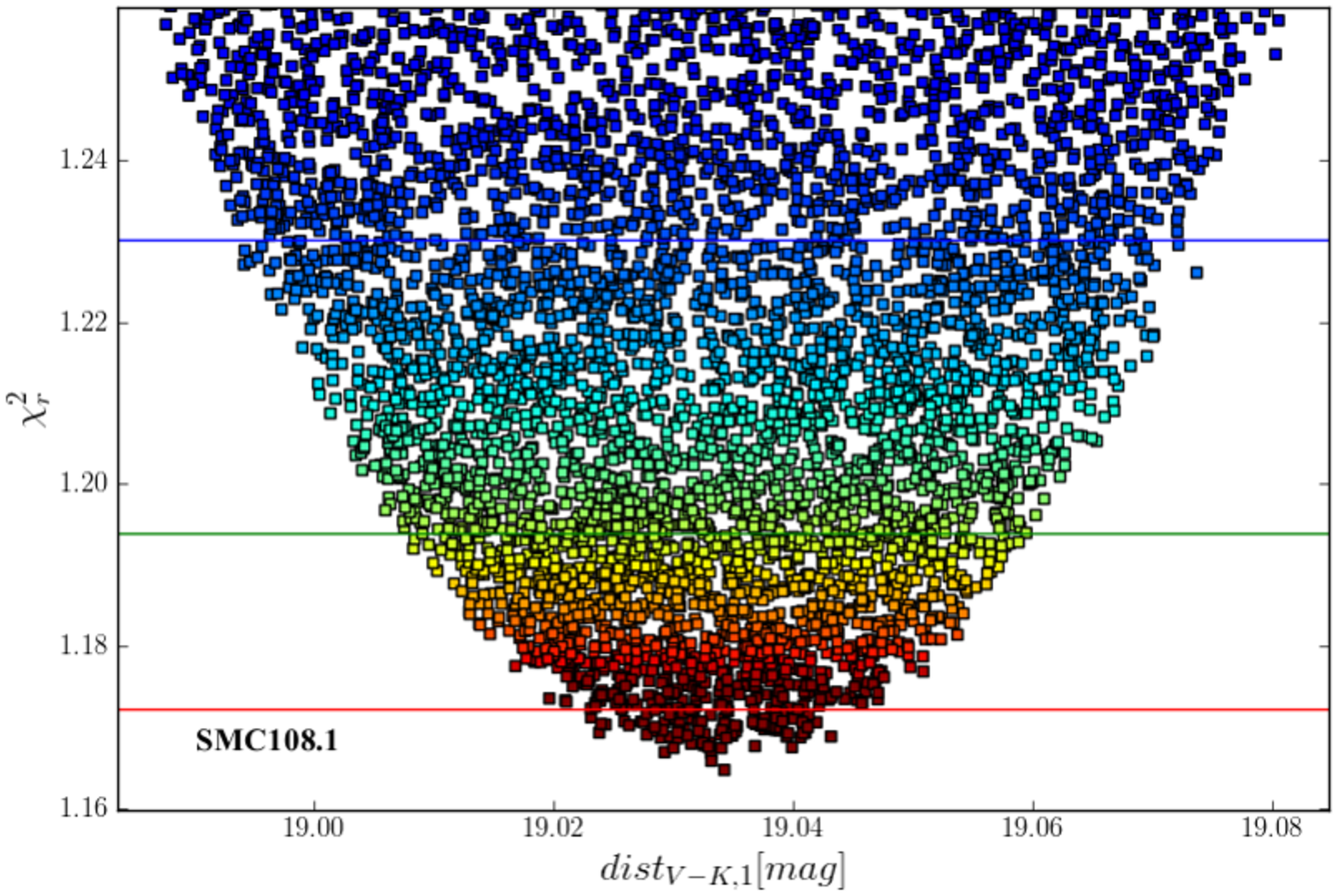}
\mbox{}
\includegraphics[angle=0,scale=.42]{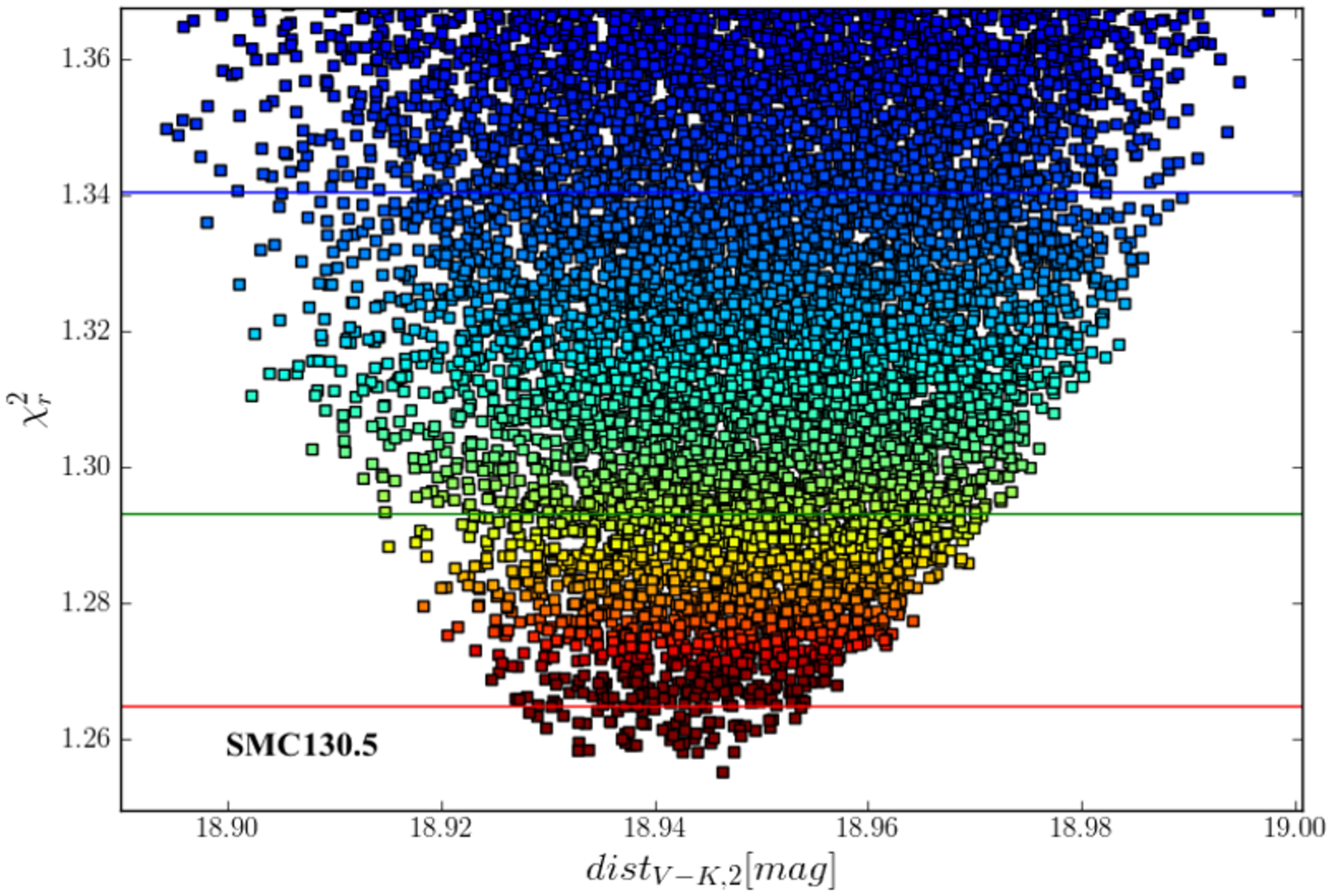}
\end{minipage}\hfill
\caption{ Reduced $\chi^2$ maps from the Monte Carlo simulations. We calculated about 30 thousand models for each eclipsing binary using the WD code to investigate the dependence of the distance on the model parameters. The horizontal lines correspond to $1\sigma$, $2\sigma$ and $3\sigma$ confidence intervals in the distance modulus. \label{fig11}}
\end{figure*}

\subsection{SMC101.8 14077}
This variable star is the faintest and bluest  analyzed in this paper. The best solution for this system was obtained with a  third light contribution included. If we assume that the presence of the third light  is real and is not connected with some imperfections in the absolute calibration of the OGLE photometric data then this  suggests the presence of a red companion or a blended star with an almost negligible contribution to the optical portion of spectrum. Indeed we could not detect any additional source of absorption lines in our spectra. For the  purpose of the determination of the distance we assumed that the third light contribution in the K-band is at least equivalent to that in the I-band. The uncertainty in the amount of third light in the K-band is the largest source of statistical error of the absolute stellar radii and the distance determination to the system.

This eclipsing binary has the identifier SC4 192903 in OGLE-II database. It was previously analyzed by \cite{gra03} who used only the photometric data  from the second phase of the OGLE project \citep{uda98} without any spectroscopic observations. A comparison of the photometric parameters presented in our Table~\ref{tbl-3} and in Table~2 from \cite{gra03}  shows very similar model solutions with the only difference resulting from our inclusion of a contribution from  third light. This results in  different measured orbital inclinations. Although the stellar mass estimate in \cite{gra03} seems to be too low (see  Tab.~3 in that paper), the stellar radii, temperatures and luminosities show good agreement.  
     
\subsection{SMC108.1 14904}
The system contains the most luminous and massive late type giants in the Magellanic Clouds analyzed by our team. The system was analyzed before by \cite{gra03} using data from OGLE-II. As in the case of SMC101.8 14077, our photometric solution is almost identical with that reported by \cite{gra03}, but a difference appears in the absolute dimensions which are significantly underestimated in the previous study. The results of our preliminary work on this system, reported by \cite{gra13}, are not significantly different from those reported in this paper, and based on more extensive  observational data we refine the previous distance estimate to this binary.  

An interesting feature of this system is the difference of the center-of-mass radial velocities between the components. The hotter primary has systematically blueshifted radial velocities with  respect to the secondary star amounting to 0.78 km s$^{-1}$ - see Fig.~3 in \cite{gra13}. We accounted for this effect by subtracting the difference and solving the radial velocities equating the systemic velocity of the secondary with the barycenter velocity of the system.  Some possible reasons for the difference are discussed in \cite{tor09} for Capella, a physically similar but lower mass binary system.

Another irregularity is a discrepancy between the temperature ratio $T_2/T_1$ as obtained from the analysis of the atmospheric parameters (see Section~\ref{atmo} and Table~\ref{tbl-4})  compared to that arising from the  light curve analysis (Table~\ref{tbl-3}). The light curve solution and the spectroscopic light ratios suggest a temperature difference of about $\Delta T=720$~K, while the atmospheric analysis suggests a difference of $\Delta T=150$~K. We repeated the atmospheric analysis by fixing the gravities of both stars to reliable values obtained from the dynamical solution reported in Table~\ref{tbl-6}. However, we obtained the same small difference of the effective temperatures. We also performed additional analysis by forcing the same metallicity for both components. However, the difference in effective temperatures do not vary again. Probably  the problem with this system is the low S/N of the spectrum of the secondary star. While the primary has a nice spectrum and by consequence the parameters are well established, for the secondary both $T_{eff}$ and $v_t$ are uncertain, and by consequence the metallicity. It is worth noting that the mean temperature  of the system $(T_1 + T_2)/2=5275$~K obtained from the atmospheric analysis is the same as the mean temperature calculated from color calibrations (i.e. 5315~K) to within the errors.    

\subsection{SMC126.1 210}
This system contains two stars with very similar masses and surface temperatures but with quite different radii. It is the only eccentric system in our sample. The system is very well-detached with no proximity effects visible in the light curve, not surprising given its long orbital period.  The best solution is found for a logarithmic law of limb darkening, a marginal improvement to that found using a linear law of limb darkening. 

An analysis of the broadening of the absorption lines provides an estimate of the ratio of the projected rotational velocities, $v_1/v_2\approx1.1$. This value is consistent with the ratio of the radii and signifies similar periods of rotation for the components (tidal locking of both stars),  in  spite of their relatively large separation.   

\subsection{SMC130.5 4296}
This is only system in our sample  with total eclipses. This allows us to determine the stellar radii with a precision of better than 1\% for both components. This eclipsing binary shows quite large proximity effects because of the relatively large secondary component. An interesting feature is the reversed luminosity ratio of the components: the less massive secondary seems to be more evolved and much brighter than the hotter primary component. Taking into account the relative proximity of the stars, we cannot exclude that there has been  an episode of mass transfer in the system when the present secondary star filled its Roche-lobe as it evolved along the red giant branch. The present system is detached, and such past mass exchange does not influence our distance determination.  
 
\section{Distance determination}
\label{distan}

\begin{deluxetable*}{lccccccccc}
\tabletypesize{\scriptsize}
\tablecaption{The distance moduli \label{tbl-7}}
\tablewidth{0pt}
\tablehead{
\colhead{ID} & \colhead{$(m\!-\!M)$} & \colhead{$\sigma{\rm A}$}&\colhead{$\sigma({\rm MonteCarlo})$}&\colhead{$\sigma E(B\!-\!V$)} &\colhead{$\sigma V$}&\colhead{$\sigma K$}& \colhead{$\sigma l_{3,K}$}&\colhead{Combined error} & \colhead{Distance} \\
& \colhead{(mag)} &\colhead{(mag)} & \colhead{(mag)}&\colhead{(mag)} &\colhead{(mag)}&\colhead{(mag)}& \colhead{(mag)} &\colhead{(mag)}& \colhead{(kpc)} 
}
\startdata
SMC101.8  14077& 19.057 & 0.006 &0.021& 0.012 & 0.001& 0.005& 0.042&0.049 & $64.79 \pm 1.48$\\
SMC108.1 14904 & 19.032 & 0.004&0.014& 0.011& 0.001& 0.004& -&0.019 & $64.03 \pm 0.55$ \\
SMC126.1 210 & 18.979 & 0.009&0.022& 0.008& 0.001& 0.004& -&0.025 & $62.49 \pm 0.74$\\
SMC130.5 4296 & 18.948 & 0.006&0.018& 0.010& 0.002& 0.009& -&0.023 & $61.60 \pm 0.67$ 
\enddata
\tablecomments{Distance moduli determinations to individual targets together with an  error budget. The errors quoted are only statistical uncertainties. Each measurement has additional an systematic uncertainty equal to 0.048 mag. }
\end{deluxetable*}

Distance estimates were derived by employing a calibration of the relation between $V$-band surface brightness  and the \mbox{$V\!-\!K$} color by \cite{ben05}. This relation (SBR) was derived from precision measurements of the angular diameters of a sample of giant and dwarf stars using  interferometry. The individual distances were calculated according to equations 4 and 5 in \cite{gra12} and the results are summarized in Table~\ref{tbl-7}. The resulting distance modulus is the average  of the measured distances of each star of an eclipsing binary. The differences in distance moduli between the components of the same system are very small. The largest discrepancy is 0.003 mag in the case of SMC101.8 14077. The unweighted mean distance modulus from our four systems is (m$-$M)$=19.004$ mag with a dispersion of only 0.050 mag. The new systems clump quite tightly around this value. 

To derive the distance to the main body of the SMC galaxy we combined the present estimates with the distance determination to another late type eclipsing binary published by our team \citep{gra12}.  Figure~\ref{fig8} shows all our estimates. The system previously analyzed - SMC113.3 4007 - is the closest one. It lies in the north-east part of the SMC (see Fig.~\ref{fig0}) which is, in fact, reported as being closer to Earth than the main body of the galaxy by a number of studies \citep{mat11,sub12,has12}. The unweighted and weighted mean distance moduli from the five systems are different by $\Delta m=0.023$ mag (0.66 kpc), with the former being longer. The weighted mean is dominated by the estimate to SMC113.3 4007 which has the smallest statistical uncertainty but is significantly off of the rest of our sample. A final distance is estimated as follows. We calculated the weighted mean distance modulus from four new binaries as $(m\!-\!M)=18.998 \pm 0.008$ mag. We then combined this result with the fifth system, obtaining $\mu_{SMC}=18.965 \pm 0.025$ mag. The  statistical uncertainty of this final estimate  arises by  combining  the standard deviation of the mean (0.008 mag) and the difference $\Delta m$ in quadrature. The resulting uncertainty  is dominated by the uncertain structure of the SMC and possible selection effects. The total uncertainty must include an  additional systematic error of 0.048 mag (Section~\ref{error}). Our final  distance estimate to the SMC is $62.1 \pm 2.0$ kpc. 

\begin{deluxetable}{@{}lcc}
\tabletypesize{\scriptsize}
\tablecaption{Distance modulus differences \label{tbl-8}}
\tablewidth{0pt}
\tablehead{
\colhead{Method} & \colhead{$\Delta\mu$ (mag) }  & \colhead{Reference} 
}
\startdata
Tip of RGB & $0.44 \pm 0.05$ &  \cite{cio00}  \\
Cepheids & $0.50 \pm 0.10$ & \cite{gro00} \\
Quattuor\tablenotemark{a}&$0.50 \pm 0.05$ & \cite{uda00}\\
Red Clump & $0.47 \pm 0.02$ & \cite{pie03} \\
Tip of RGB & $0.50 \pm 0.03$ & \cite{pie03} \\
Tip of RGB & $0.40 \pm 0.12$ & \cite{sak04}\\
RR Lyrae & $0.39 \pm 0.04$&  \cite{sze09} \\
Cepheids &$0.44 \pm 0.12$& \cite{bon10} \\
Red Pulsators & $0.41 \pm 0.02$ & \cite{tab10} \\
Type II Cepheids & $0.39 \pm 0.05$ & \cite{mat11}\\
Cepheids & $0.43 \pm 0.05$ & \cite{mat11} \\
IRSB Cepheids & $0.47 \pm 0.15$ & \cite{sto11}\\
RR Lyr &  $0.61 \pm 0.20 $ & \hspace*{-0.4cm} \cite{kop12} \\
IRSB Cepheids & $0.44 \pm 0.06$ & \cite{gro13} \\
FU Cepheids & $0.48 \pm 0.03$ & \cite{ino13} \\
FO Cepheids & $0.52 \pm 0.03$ & \cite{ino13} \\
Eclipsing Binaries\tablenotemark{b} & $0.472 \pm 0.026$ & this paper
\enddata
\tablenotetext{a}{Four methods: average of Red Clump, Tip of RGB, RR Lyr and classical Cepheids}
\tablenotetext{b}{Late type systems.}
\tablecomments{Recent distance moduli differences between Magellanic Clouds based on different methods.}
\end{deluxetable}

\begin{figure}
\includegraphics[angle=0,scale=.50]{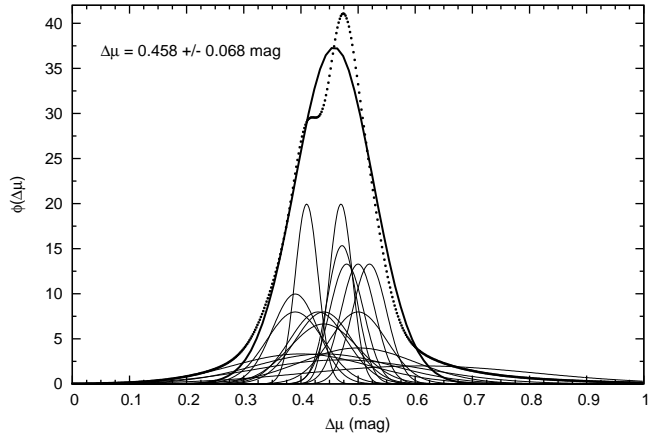}
\caption{ The normal distributions of the relative distance moduli between Magellanic Clouds $\Delta\mu\!=\!\mu_{SMC}-\mu_{LMC}$ taken from Table~\ref{tbl-8} - thin lines. The distribution of their sum is shown as points and it is fitted by a single normal distribution, plotted as the thick line. \label{fig10}}
\end{figure}

\subsection{Error Budget}
\label{error}
The main contributions to the statistical uncertainties for all systems are presented in Tab.~\ref{tbl-7} (columns 3-7). The error in the semimajor axis is given as $\sigma{\rm A}$ and an uncertainty of 1\% in the absolute scale of the system translates to a 0.022 mag error in the distance modulus. We performed Monte Carlo simulations as described in \cite{pie13} for each of our binaries - see Fig.~\ref{fig11}. The resulting uncertainty of the distance is reported in  column 4. This uncertainty takes into account the error in the determination of the relative radii, magnitude disentangling, and correlations between model parameters in the WD code. The main contribution to the uncertainty returned by the Monte Carlo simulations is the error in the sum of the relative radii $r_1\!+r_2$. The statistical uncertainty in the extinction was assumed to be 0.020 mag in each case. Columns 6 and 7 give the standard deviations of the mean brightness at maximum light in the V- and K-band, respectively. Remaining sources of statistical uncertainty, i.e.: model atmosphere approximations, disentangling of individual magnitudes, the adopted limb darkening law, contribute at a level of  0.001 mag \citep{gra12,pie13}, and these are insignificant in the total statistical error budget.     

The sources of systematic uncertainty in our method are the following: the uncertainty of the empirical calibration of surface brightnesses by \cite{ben05} (0.040 mag), uncertainty in the extinction (0.020 mag, translating to a 0.010 mag error in distance), metallicity dependence on the surface brightness - color relation (0.004 mag) and zero point errors of V- and K-band photometry (0.010 mag and 0.015 mag, respectively). Combining these in quadrature we obtain a systematic error of 0.048 mag  in the distance modulus of each eclipsing binary.   

\section{Discussion}
 
 \subsection{Relative distance between the Magellanic Clouds}
 \label{relativ}
 \cite{san68} used observations of classical Cepheids to  unambiguously show that the Magellanic Clouds have different distance moduli, corresponding to a distance difference of about 11 kpc. This difference is small enough that in practice almost all methods of distance determination can be employed simultaneously to both objects.  In fact the relative distance between the galaxies obtained from the application of a particular method is usually better constrained than the individual distances determined by that method. This is because distance determination methods are dominated by systematic uncertainties which mostly cancel out when used to obtain the distance to each galaxy with the same observational setup and similar quality data. 
 
In Table~\ref{tbl-8} we summarize recent determinations of distance moduli differences between the Magellanic Clouds ($\Delta\mu\!=\!\mu_{SMC}-\mu_{LMC}$) by authors who used exactly the same method for both galaxies. Distance moduli based on old population stars are metallicity corrected. In the case of our method we adopted the distance modulus  to the LMC from \cite{pie13}, $\mu_{LMC}=18.493\pm0.006$ mag. The interpretation of Table~\ref{tbl-8} is hindered by the fact that some authors do not clearly separate  statistical and systematic uncertainties. The average from the young "metal rich" population containing classical Cepheids and eclipsing binary systems is  $\Delta\mu=0.472$ mag and the average from the old "metal poor" population is $\Delta\mu=0.453$ mag. We summarize these distributions in  Fig.~\ref{fig10}. The resulting distribution is slightly bimodal with a main peak at the position of the young population average. A gaussian fit to the distribution gives $\Delta\mu=0.458 \pm 0.068$ mag. 
 
\subsection{The geometrical depth of the SMC}
The extended structure of the SMC in the line of sight was suggested  by \cite{jon61}. Over the last 30 years  substantial evidence has been accumulated about the intricate and extended structure of the SMC from the studies of early type stars and HII regions \citep{mar89}, stellar clusters \citep{cro01}, classical Cepheids \citep{wel87,gro00,has12}, RR Lyr stars \citep{sub12,has12}, red clump stars \citep{gar91,hat93,sub09,sub12}, eclipsing binary stars \citep{hil05,nor10} and red giant branch stars \citep{lah05}. The north-east part of the SMC was shown to be closer to us and have a substantial degree of complexity, and it is reported also to have the largest geometrical depth in line-of-sight \citep{sub12}. The apparent difference in the distance to the  eclipsing binary SMC113.3 compared to the rest of our sample may confirm the existence of  substructure in the front of the galaxy. We can interpret that the remaining eclipsing binaries  are part of the main body of the galaxy.

The small dispersion in the distance moduli of our four binaries (0.05 mag, corresponding to $\sim1.4$ kpc) may signify a small geometrical depth of the main part of the SMC. We can investigate this more by an analysis of the distribution of differences in the distance moduli  (Fig.~\ref{fig10}). The cumulative distribution has a dispersion of 0.068 mag. Because most of the systematic effects do not influence  $\Delta\mu$, and the geometrical depth of the central part of the LMC is small \citep{pie13}, we can interpret this spread as an imprint of the geometrical depth of the main part of the SMC. At a distance of 62 kpc this dispersion corresponds to a 1-$\sigma$ line-of-sight depth of 4 kpc. \cite{has12} gave a review of recent line-of-sight depth estimates for the SMC which we can use for a comparison. They concluded that  the  1-$\sigma$ line-of-sight depth is between 4 and 5 kpc  for the old population of stars (RR Lyr and red clump stars). For  the Cepheids the estimated depth is larger by up to $\sim 8$ kpc, and in the case of intermediate-age clusters the depth reaches $\sim 10$ kpc.       

\subsection{Distance to the SMC}
\cite{gra12,gra13} compiled some recent distance determinations to the SMC as reported  up to 2011. Their conclusion was that the canonical value of the distance modulus to the SMC ($\mu_{SMC}=18.90$ mag, \citep{wes97}) remained marginally consistent with most recent determinations (see also Section~\ref{relativ}). However, modern determinations prefer a  slightly larger distance moduli of $\mu_{SMC}=18.95 \pm 0.07$ mag.  The inclusion of the most recent distance determinations \citep{has12,gro13,ino13} does not affect this finding. In fact, if we adopt  our estimate of the relative distance modulus between the Magellanic Clouds and we assume $\mu_{LMC}=18.493$ mag,  we derive $\mu_{SMC}=18.493+0.458=18.951$ mag. This result remains in perfect agreement with our distance modulus measured with late type eclipsing binaries, and being based on a number of independent standard candles, we advocate here its adoption as the "canonical" distance modulus to the Small Magellanic Cloud. 

\acknowledgments
We are gratefull for financial support from Polish National Science Center grant MAESTRO 2012/06/A/ST9/00269 and the TEAM subsidy from the Foundation for Polish Science (FNP). Support from the BASAL Centro de Astrof{\'i}sica y Tecnolog{\'i}as Afines (CATA) PFB-06/2007 is also acknowledged. Based on observations made with ESO 3.6m and NTT telescopes in La Silla under programme 074.D-0318, 074.D-0505, 082.D-0499, 083.D-0549, 084.D-0591,  086.D-0078, 091,D-0469(A) and CNTAC  programme CN2010B-060. We also thank  the staffs at La Silla Observatory (ESO) and Las Campanas Observatory (Carnegie) for their excellent support and the anonymous referee who helped us to improve this paper.  

The OGLE project has received funding from the European Research Council  under the European Community's Seventh Framework Programme  (FP7/2007-2013)/ERC grant agreement no. 246678 to AU. 

WG gratefully acknowledges the hospitality and support of Professor  Ralf Bender during his sabbatical leave at Munich Sternwarte and the Max Planck Institute for Extraterrestrial Physics in Garching.
{}    

\end{document}